\documentclass[prl,twocolumn,amsmath,amssymb,superscriptaddress]{revtex4-1}

\usepackage{graphicx,picture,calc}
\usepackage{verbatim}
\usepackage{braket}
\usepackage{epsfig}
\usepackage{epstopdf}
\usepackage{amsfonts}
\usepackage{amsthm}
\usepackage{amsmath}
\usepackage{amssymb}
\usepackage{color}
\usepackage[usenames,dvipsnames,svgnames,table]{xcolor}
\usepackage{hyperref} 
\hypersetup{colorlinks=true,linkcolor=NavyBlue,citecolor=BrickRed,urlcolor=NavyBlue}
\usepackage{dsfont}
\usepackage{color}
\usepackage{grffile}
\usepackage{bm}

\begin{document}

\title{A Striped Electron Fluid on (111) KTaO$_3$}
\author{P. Villar Arribi}
\affiliation{Materials Science Division, Argonne National Laboratory, Lemont, IL  60439, USA}
\author{A. Paramekanti}
\affiliation{Department of Physics, University of Toronto, Toronto, Ontario M5S1A7, Canada.}
\author{M. R. Norman}
\affiliation{Materials Science Division, Argonne National Laboratory, Lemont, IL  60439, USA}

\date{\today}

\begin{abstract}
A recent study has revealed that the low carrier density electron gas (2DEG)  induced at the interface of EuO
and (111) KTaO$_3$ exhibits a broken symmetry phase with a strong in-plane anisotropy of the resistivity.
We present a minimal tight binding model of this (111) 2DEG, including the large spin-orbit coupling
from the Ta ions, which reveals a hexagonal Fermi surface with a highly enhanced 2$k_F$ electronic susceptibility.
We argue that repulsive electronic interactions, together with a ferromagnetic EuO substrate, favor a magnetic stripe instability leading to a partially gapped Fermi surface.
Such a stripe state, or its vestigial nematicity, could explain the observed transport anisotropy. We propose a $k\cdot p$ theory for the low energy
$j=3/2$ states, which captures the key results from our tight-binding study, and further reveals the intertwined dipolar and octupolar modulations underlying this
magnetic stripe order. We conclude by speculating on the relation of this stripe order to the superconductivity seen in this material.
\end{abstract}


\maketitle

The superconducting 2DEG at the surface of SrTiO$_3$ has been the subject of much investigation since
its observation back in 2007 \cite{reyren}.  A few years later, its $5d$ analog, KTaO$_3$ (KTO), was found to exhibit superconductivity (SC) at $50$\,mK
in an (001) 2DEG created using ionic liquid gating \cite{ueno}. 
A recent experiment discovered that for (111) oriented KTO, $T_c$ is dramatically enhanced, by a factor of 40, with SC
occurring up to $\sim\! 2$\,K at carrier densities  $n \sim\! 10^{14}$/cm$^2$ \cite{liu}.  Even more remarkably, at low carrier densities where SC
occurs with $T_c \!\sim\!0.5$\,K, it descends from an apparent nematic phase with a significant in-plane resistance anisotropy of $\sim\!3$ \cite{liu}.  This anisotropy onsets abruptly
at a higher temperature $2.2$\,K, suggesting a phase transition into an ordered state.  At zero magnetic field, the anisotropy is only
observed if KTO is in contact with EuO (for KTO on LaAlO$_3$, an in-plane magnetic field is required for its
observation \cite{liu}).  As EuO is ferromagnetic, magnetism is likely to
play an important role in this phenomenon.  The presence of charge, spin, and superconducting correlations as a 
function of carrier concentration for (111) KTO is reminiscent of a number of other materials such as cuprates~\cite{cuprates_nem_1,cuprates_nem_2}, 
iron pnictides and chalcogenides~\cite{Paglione_IBSC}, doped  Bi$_2$Se$_3$~\cite{bi2se3_1,bi2se3_2,bi2se3_3}, and twisted bilayer graphene 
near a magic angle~\cite{TBG_nematic}. Furthermore, since the conduction band of KTO arises from 
spin-orbit coupled (SOC) $j\!=\!3/2$ states \cite{Matthias_1972,bruno},
the broken symmetry nematic is expected to also display intertwined multipolar orders as conjectured for Cd$_2$Re$_2$O$_7$
\cite{LiangFu_PRL2015,Harter295,Norman_PRB2017,hayami2019}.

 \begin{figure}[t]
\includegraphics[width=0.98\columnwidth]{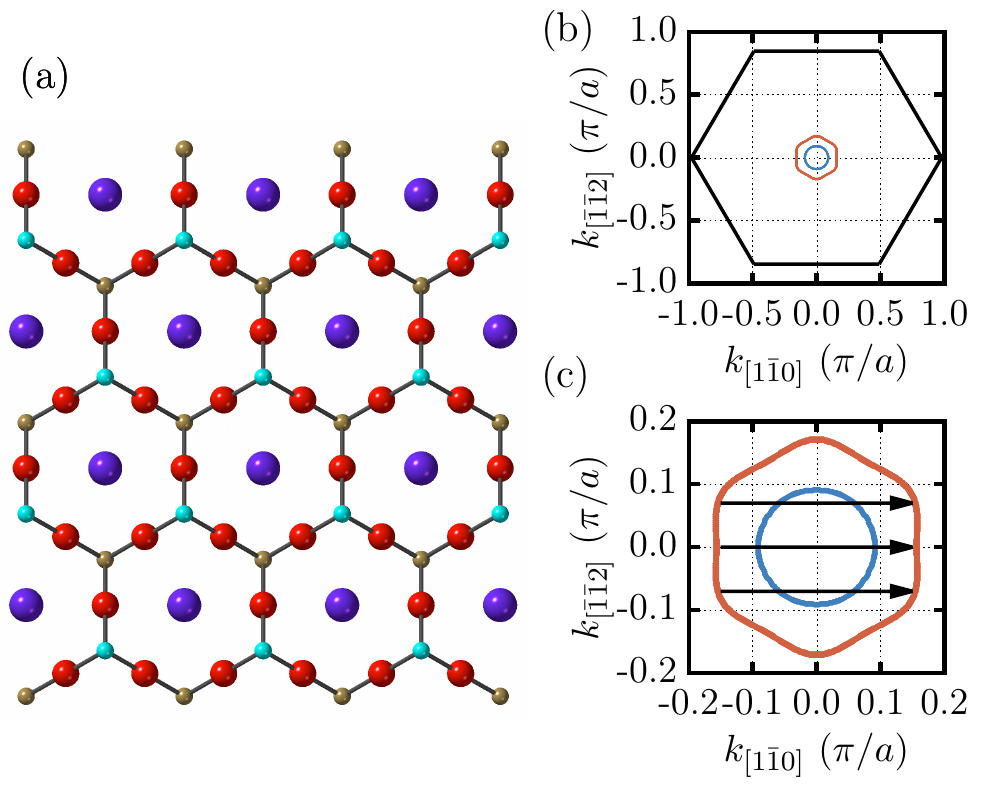}
\caption{(a) Crystal structure of the top three (111) layers of KTO (Ta, KO$_3$, Ta), with the horizontal axis along (1,-1,0) 
and the vertical axis along (1,1,-2).  K ions are in purple, O ions in red, Ta ions in the top layer in cyan and Ta ions in
the bottom layer in gold. (b) Fermi surface of (111) KTO from the bilayer tight-binding model and the carrier density of the
nematic state. (c) Blow up of (b), showing the nesting wavevectors associated with the outer sheet of the Fermi surface.
$a$ refers to the cubic lattice constant of KTO.}
\label{fig1}
\end{figure}

KTO is a band insulator with a large gap $\sim3.6$\,eV \cite{Jellison_PRB2006}. 
Experiments have realized both a (001) 2DEG \cite{Nakamura_PRB2009,King_PRL2012,Rozenberg_PRB2012,Sung_ACSNano2019,liu}
and a (111) 2DEG \cite{bareille,bruno,liu} at the free surface of KTO, due to oxygen vacancies induced by cleaving or by irradiating the surface,
as well as at KTO interfaces with oxides such as LaAlO$_3$ and EuO. Fig.~\ref{fig1} shows the crystal structure of (111) KTO, consisting
of alternating layers of Ta and KO$_3$, with each Ta layer forming a triangular lattice. This structure
is highly polar given the $5+$ nature of the Ta ions. Angle resolved photoemission spectroscopy (ARPES)  of the 2DEG revealed
six-fold symmetric Fermi surfaces (FSs) \cite{bareille,bruno}. The observed bands were found to be captured by a (111)
bilayer model \cite{bareille}, consisting of $t_{2g}$ orbitals from two Ta layers forming a buckled honeycomb plane,
a setting proposed for realizing topological phases by Xiao {\it et al.}~\cite{xiao}.

We find that a minimal two-parameter honeycomb tight-binding model provides a good
description of the ARPES observations; details can be found in  \cite{suppmat} and
Refs.~\cite{xiao,khalsa,bruno,liu}.
The dominant scale is the Ta-O-Ta hopping, $t \sim 1$\,eV, which is proportional to $t_{pd}^2/\Delta_{pd}$ where $t_{pd}$ is 
the hopping between the Ta $5d$ $t_{2g}$ and O $2p$ orbitals, and $\Delta_{pd}$ is the Ta-O charge transfer energy. Since
each Ta $t_{2g}$ orbital hops to only two of the three nearest neighbors on the honeycomb lattice, the resulting bands exhibit 1D character.
The FS of each orbital forms a spin-degenerate pair of parallel lines, with the FS of different orbitals rotated with respect to 
each other by 120 degrees~\cite{suppmat}.  The inclusion of SOC, $\lambda \!\sim\! 0.25$\,eV,
which is the next largest energy scale, dramatically
alters the electronic structure. SOC hybridizes the different 1D dispersions, and splits the $t_{2g}$ manifold at the $\Gamma$-point
into a lower $j\!=\!3/2$ quartet and an upper $j\!=\!1/2$ doublet, separated by $\sim\!0.4$\,eV.
The net result is that the parallel FS lines break up and reconstruct into closed FSs. 
Motivated by Ref.~\onlinecite{bruno}, we supplement this minimal model with a small trigonal distortion term so
the $j\!=\!3/2$ quartet at $\Gamma$ splits into two Kramers doublets separated by $\sim\!15$\,meV.
Our model provides a
reasonable description of the recent ARPES data \cite{bruno,suppmat}, at a density $n\! \sim\! 10^{14}$/cm$^2$,
which reveals a larger outer star-shaped FS along with an inner 
hexagonal FS, both centered at $\Gamma$ \cite{1supp}.

The recent experiments which observe nematic transport correspond to lower
densities $n \!\sim\! 3.5 \! \times\! 10^{13}$/cm$^2$. To explore this regime, we start from our well-motivated
model above, and lower the chemical potential to achieve this density. The resulting 
FS, shown in Fig.~\ref{fig1}, reveals an inner circular FS, and an outer hexagonal
FS which has flat faces nested along the $\Gamma$-$K$ directions. We examine below the consequences of this 
nesting for $2k_F$ stripe order and transport anisotropy. The FSs shown in Fig.~\ref{fig1} are spin-degenerate;
we later incorporate weak Rashba spin splitting \cite{bareille,bruno} induced by the broken inversion symmetry at the interface.

\begin{figure}
\includegraphics[width=0.99\columnwidth]{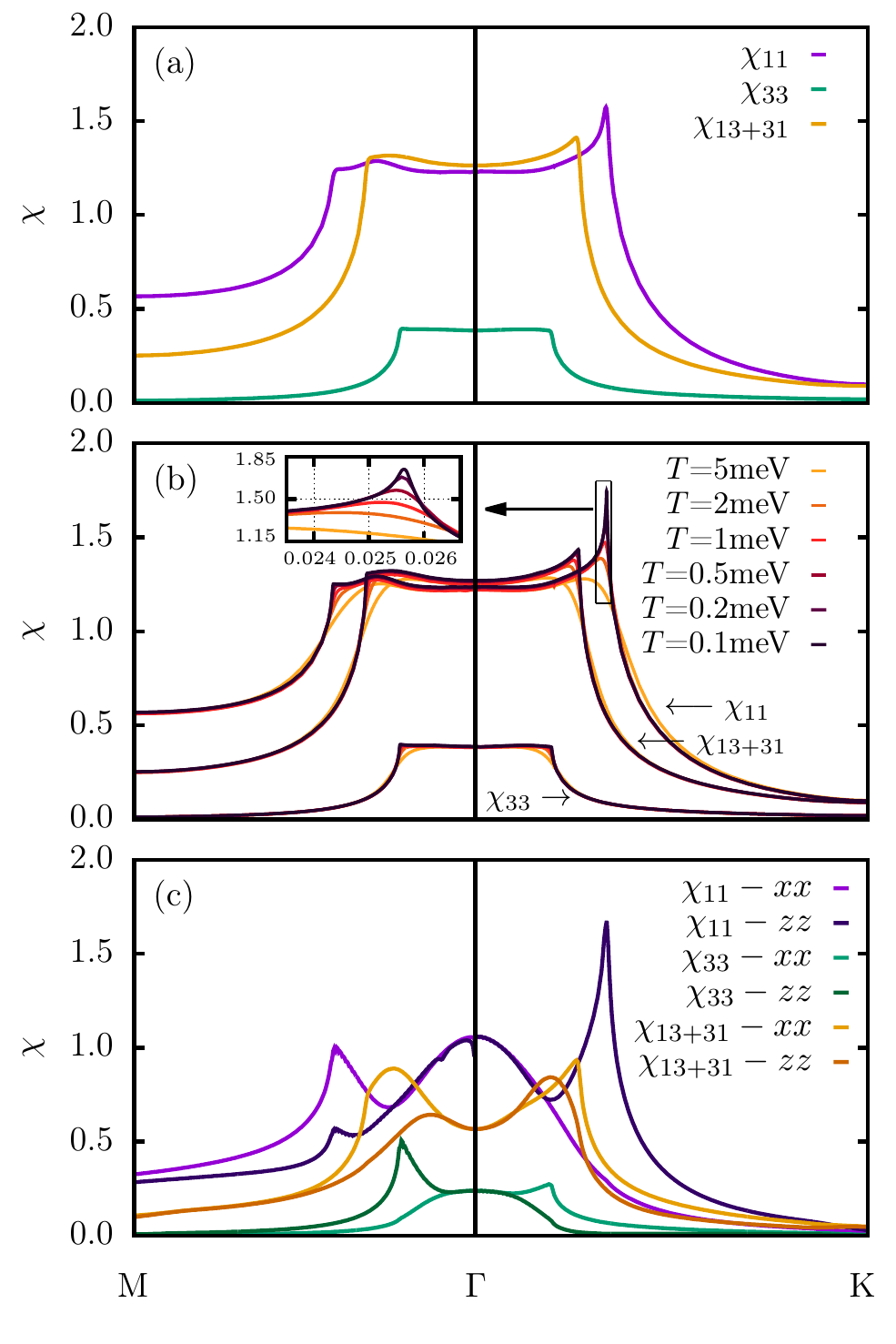}
\caption{(a) Lindhard susceptibility, $\chi_0$, along $\Gamma$-$M$ and $\Gamma$-$K$ 
computed with $T=0.5$~meV. Subscripts $1,3$ are band indices ($1$ for the outer FS, $3$ for the inner FS,
noting that bands 1,2 and 3,4 are Kramers degenerate).
(b) Evolution of the susceptibility with temperature. Inset: detail of the cusp in $\chi_{11}$ along the 
$\Gamma$-$K$ direction associated with the outer FS. (c) Lindhard susceptibility including spin matrix elements as defined in Eq.~\ref{eq:mat_el}. 
$x$ and $z$ correspond to the spin operators $S_x$ and $S_z$.
The strong (11) cusp is only found for the $zz$ component.}
\label{fig2}
\end{figure}

\newcommand{\bk}{{\bf k}}
\newcommand{\bq}{{\bf q}}
\newcommand{\bQ}{{\bf Q}}

The susceptibility for such a hexagonal FS is expected to resemble the 1D Lindhard function which diverges 
logarithmically in $T$ at the nesting wavevector. Interestingly, the nesting direction is along  $\Gamma$-$K$ (i.e., the (1,-1,0) direction) 
which corresponds precisely to the observed high resistivity direction in the nematic phase \cite{liu}. In a stripe model for
the nematic phase, one would indeed anticipate that the resistivity is maximal along the stripe wavevector, $\bq_s$.
To investigate this further, we calculate the Lindhard susceptibility, $\chi_0$, for the bilayer model:
\begin{equation}
\chi_0(\bq) = \sum_{\bk,n,m} \frac{f_{\bk,n} - f_{\bk+\bq,m}}{\epsilon_{\bk+\bq,m}-\epsilon_{\bk,n}+i\delta}
\end{equation}
where $m$ and $n$ are band indices, $f$ are Fermi functions, $\epsilon$ are the band energies,
and $\delta$ is a small broadening (set to 0.1 meV or smaller).
Although the bilayer model has six spin-degenerate pairs of bands (three $t_{2g}$ orbitals, two layers), only the lowest two are relevant at low energy
and we confine our discussion to them. To begin with, we will be agnostic concerning spin versus charge, and therefore not include
matrix elements until later.  The resulting $\chi_0$, decomposed in terms of $n$ and $m$, is shown in 
Fig.~\ref{fig2}a. As expected, the outer hexagon gives rise to a susceptibility maximum along $\Gamma$-$K$ 
due to nesting of each of the two parallel sides of the hexagon (as indicated in Fig.~\ref{fig1}c). This is evident from the 
cusp-like behavior of the intraband $\chi_0$, indicating quasi-1D behavior. Fig.~\ref{fig2}b shows
that this cusp becomes better defined upon lowering $T$, as expected.

To proceed further, we need to consider matrix elements.  In the experiments, the nematic phase at zero magnetic 
field is found at the KTO-EuO interface, but not at the KTO-LaAlO$_3$ interface. This indicates that magnetism is playing a key role.
This can be understood from the fact that the Eu $4f$ electrons exhibit ferromagnetic order with a large moment.
These $4f$ electrons overlap with the Eu $5d$ orbitals which in turn overlap with the Ta $5d$ electrons (the Ta to oxide layer 
spacing in (111) KTO is only 1.15~\AA{}, whereas $\langle r \rangle_{{\mathrm{Eu}}-4f} \!\sim 0.9\!$\,\AA{}, 
$\langle r \rangle_{{\mathrm{Eu}}-5d} \!\sim\! 2.7$\,\AA{}, and $\langle r \rangle_{{\mathrm{Ta}}-5d} \!\sim\! 2.2$ \AA{}~\cite{desclaux}). 
Calculations for (001) EuO-KTO find induced moments of $\sim$ 0.2 $\mu_B$ on
the first TaO$_2$ layer \cite{zhang}.
This motivates including spin matrix elements in the numerator of Eq.~1:
\begin{equation}
g^2 \bra{\bk,n}S_i(\bq)\ket{\bk+\bq,m}\bra{\bk+\bq,m}S_j(\bq)\ket{\bk,n}
\label{eq:mat_el}
\end{equation}
where $g\!=\!2$, $S_i$ are spin-$\frac{1}{2}$ operators ($i$=$x,y,z$), and $\ket{\bk,n}$ are the band eigenvectors.
Because of strong SOC, the susceptibility is anisotropic
even without the feedback from the energy gap due to density wave formation.  The results are shown in Fig.~\ref{fig2}c.
The cusp along $\Gamma$-$K$ is associated with the $zz$ component of $\chi$.  As $z$ is orthogonal to (1,-1,0), this
implies a transverse spin density wave, which is typical for a magnetic stripe model \cite{2supp}.

The mean field transition temperature is determined by the divergence of the full interacting susceptibility.
This is given by the condition $I(\bq)\chi_0(\bq,T)\!=\!1$, where $I(\bq)$ is the interaction function.
Based on the above considerations, we expect $I(\bq)$ would be induced by the combined effect of the amorphous ferromagnetic
EuO substrate and local Ta correlations, rendering it a weak function of $\bq$. Thus,
the ordering vector would be determined by the cusp in $\chi_0$.  The value of $I(\bq)$ would need to be sizable (on the 
scale of $\sim\!1$\,eV) in order to induce the transition, with the low value of $T_s$ due to the logarithmic (BCS-like) rise in the cusp
of $\chi_0$ with decreasing $T$ as can be seen in Fig.~\ref{fig2}b. In this scenario, the disappearance of nematicity 
in higher carrier density samples could be due to the reduction of $I(\bq)$ from screening, or due to enhanced disorder 
scattering as reflected by the lower mobility of such samples~\cite{liu}.

\begin{figure}
\includegraphics[width=0.55\columnwidth]{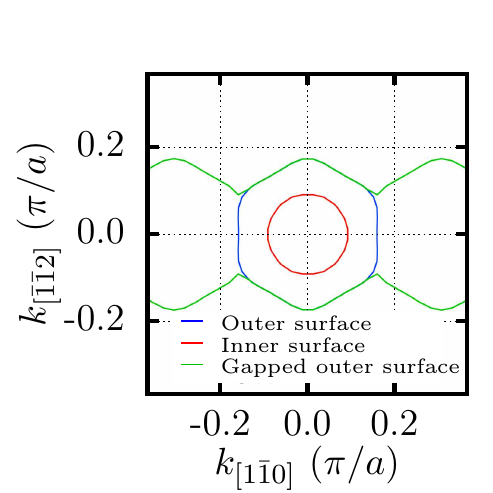}
\caption{Original FS from Fig.~\ref{fig1} (red,blue), as well as the reconstructed outer FS (green) due to
a spin density wave potential, $V(\bq_s)$, of strength $4.4$\,K, with $\bq_s\!=\!(q_s,0)$ and $q_s$ given by
the peak in $\chi_0$ along $\Gamma$-$K$ for the outer FS.}
\label{fig3}
\end{figure}

We next consider the question of transport anisotropy in this stripe state.
In Fig.~\ref{fig3}, we show the outer FS as reconstructed by a spin-density wave
within a simple calculation~\cite{arc} involving a $3\times 3$ secular matrix where one couples the states $\bk-\bq_s$
and $\bk+\bq_s$ with $\bk$, with $\bk$ from the lowest band (band $1$) and a stripe potential, $V(\bq_s)$. Here, we take
$\bq_s=(q_s,0)$, and $V(\bq_s)$ to have a typical mean-field value of 2$T_s$ where $T_s \!\sim\! 2.2$\,K from Ref.~\onlinecite{liu}. We
find that the original FS is wiped out along the nesting direction, leading to a reconstructed open FS which 
is expected to exhibit a strong resistive anisotropy. 

We briefly comment on the energetic competition between a single-$\bq$ state versus a triple-$\bq$ state. 
In the presence of SOC, for a fixed spin direction, only
one of the three equivalent $\Gamma$-$K$ directions would have a cusp, but not the other two (as in Fig.~\ref{fig2}c),
leading to transport nematicity. Based on
the above FS reconstruction, one might expect that a triple-$\bq$ state would gap out the entire outer FS (i.e., all hexagonal faces),
and thus would be energetically preferred over the single-$\bq$ one.
Such a state would not have an in-plane transport anisotropy. However, such non-coplanar
spin crystals typically arise for a non-zero perpendicular magnetization whereas the EuO magnetization is expected
to be in the plane of the interface. In this case, SOC could favor the single-$\bq$ state, with 
the moment direction locked to the stripe wavevector.
This provides further support of a magnetic stripe rather than a charge stripe order.
A complete treatment of this problem would require calculating the cubic and quartic terms in a Landau free-energy
expansion~\cite{mcmillan,ashot}, but this time including the spin matrix elements. This is an involved task, which we defer to
future work.

\begin{figure}
\includegraphics[width=0.99\columnwidth]{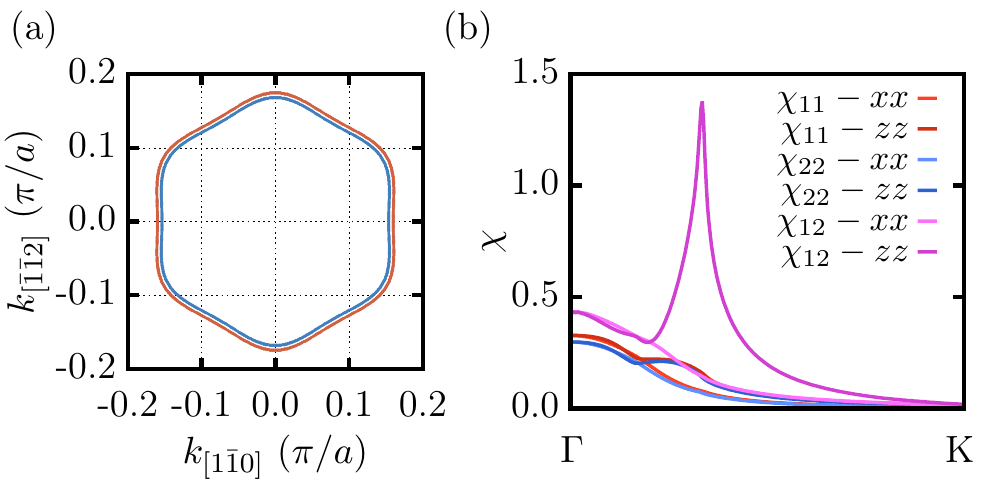}
\caption{(a) Outer FS as in Fig.~\ref{fig1}c, but including a Rashba term which lifts the
Kramers degeneracy. (b) Resulting Lindhard susceptibility, $\chi_0$, computed with $T=0.5$~meV, including spin matrix elements as defined in Eq.~\ref{eq:mat_el} plotted along the (1,-1,0) direction. 
The cusp is again associated with the $zz$ component, but now is an interband term between the two Rashba-split bands.}
\label{fig4}
\end{figure}

We next consider the effect of the Rashba spin splitting. In a single-layer model, the Rashba term
for the (111) case is given in Ref.~\onlinecite{boudjada}.  As the Ta-Ta hopping in-plane is weak, we instead
consider the Rashba term given by the Ta-O-Ta path connecting the layers, which is a (111) generalization of
the (001) case considered by Khalsa {\it et al.}~\cite{khalsa}.  Here, the largest term is due to inversion breaking
on one of the Ta-O segments followed by a $t_{pd}$ hop along the other.  The functional form is given in~\cite{suppmat}.
Both this form and the one of Ref.~\onlinecite{boudjada}, which are off-diagonal in the orbital index,
give similar results, with a relatively isotropic Rashba splitting around the FS (Fig.~\ref{fig4}a).
The effect of this is minor for $\chi_0$ without matrix elements given the small value of the Rashba splitting (of order a few meV).
However, once we include their effect, the largest $\chi_0$ contribution comes from the interband $zz$ component 
associated with the Rashba-split outer FS (Fig.~\ref{fig4}b), which can be understood from the
Rashba-induced FS spin texture.

\newcommand{\hJ}{\hat{J}}

While our tight-binding model study captures the salient observations for the (111) KTO 2DEG, it is nevertheless 
useful to construct a continuum $k\! \cdot \! p$ theory for the low energy $j\!=\!3/2$ states near the $\Gamma$-point
\cite{Luttinger_1956,LiangFu_PRX2018}.
This allows us to clearly expose the multipolar character of the magnetic stripe order. We begin with a minimal 3D Luttinger
model \cite{suppmat}, $H_{\rm bulk} \!=\!  \alpha_1\! \sum_i k^2_i  \! \hJ_i^2$, with $\alpha_1 \!=\!0.2$\,eV,
which reasonably captures the bulk band dispersion up to a momentum cutoff which we fix as $\Lambda\!=\!\pi/3$ (momenta are
in units of the inverse cubic lattice constant $1/a$).
Here, $\hJ_i$ refer to spin-$3/2$ angular momentum operators (with $i\!=\! x,y,z$), and
$\hJ_0$ denotes the $4\! \times\! 4$ identity matrix.
To describe the $(111)$ 2DEG, we project to 2D so that
\begin{eqnarray}
\!\!\!\! H^{(0)}_{2D} &\!\!=\!& \frac{\alpha_1}{6} \! \! \left[ \! (\sqrt{3} k_1 \!+\! k_2)^2\hJ^2_x  \!+\!  (\sqrt{3} k_1 \!-\! k_2)^2 \hJ^2_y \!+\! 4 k^2_2\hJ^2_z \!\right]
\end{eqnarray}
with $k_1,k_2$ being momenta respectively along the $(1\bar{1}0)$ and $(11\bar{2})$ directions.
To describe the hexagonal FS, we include the symmetry-allowed sixth-order terms,
\begin{eqnarray}
H^{(1)}_{2D} &\!=\!& \left[\beta_1 (k_+^6 \!+\! k_-^6) \!+\! \beta_2 (k_+^3\!+\! k_-^3)^2\right] \hJ_0 \notag\\
&+& \beta_3 (k_+^3 \!-\! k_-^3)^2 \hJ_3^2
\end{eqnarray}
where $\hJ_3\!=\!(\hJ_x\!+\!\hJ_y\!+\!\hJ_z)/\sqrt{3}$ and $k_\pm = k_1 \pm i k_2$.
We set 
$(\beta_1,\beta_2,\beta_3)\!=\! (0.35,0.6,-0.65)$\,eV;
although $\{\beta_i\}$ naively appear to be large energy scales, note that $\beta_i k_F^4 \! \ll \! \alpha_1$
for relevant densities. 
Finally, we incorporate two weaker 
terms: a trigonal distortion $\tilde{\Delta}$, and a Rashba coupling $\tilde{\gamma}$ from inversion breaking,
\begin{eqnarray}
\!H^{(2)}_{2D} &\!=\!&\tilde{\Delta} \hJ_3^2  + \tilde{\gamma} (\hJ_1 k_2 \!-\! \hJ_2 k_1)
\end{eqnarray}
where $J_1 \!=\! \frac{1}{\sqrt{2}}(J_x\!-\! J_y)$, $J_2 \!=\! \frac{1}{\sqrt{6}} (J_x\!+\!J_y\!-\!2 J_z)$,
and we fix $(\tilde{\Delta},\tilde{\gamma}) \!=\! (7,7)$\,meV. Our $k\cdot p$ model for the (111) 2DEG,
$H_{\rm 2DEG} \!= \!H^{(0)}_{2D}+\!H^{(1)}_{2D}+\!H^{(2)}_{2D}$, reasonably reproduces the tight-binding FSs from low to moderate densities \cite{suppmat}.

\begin{figure}
\includegraphics[width=0.99\columnwidth]{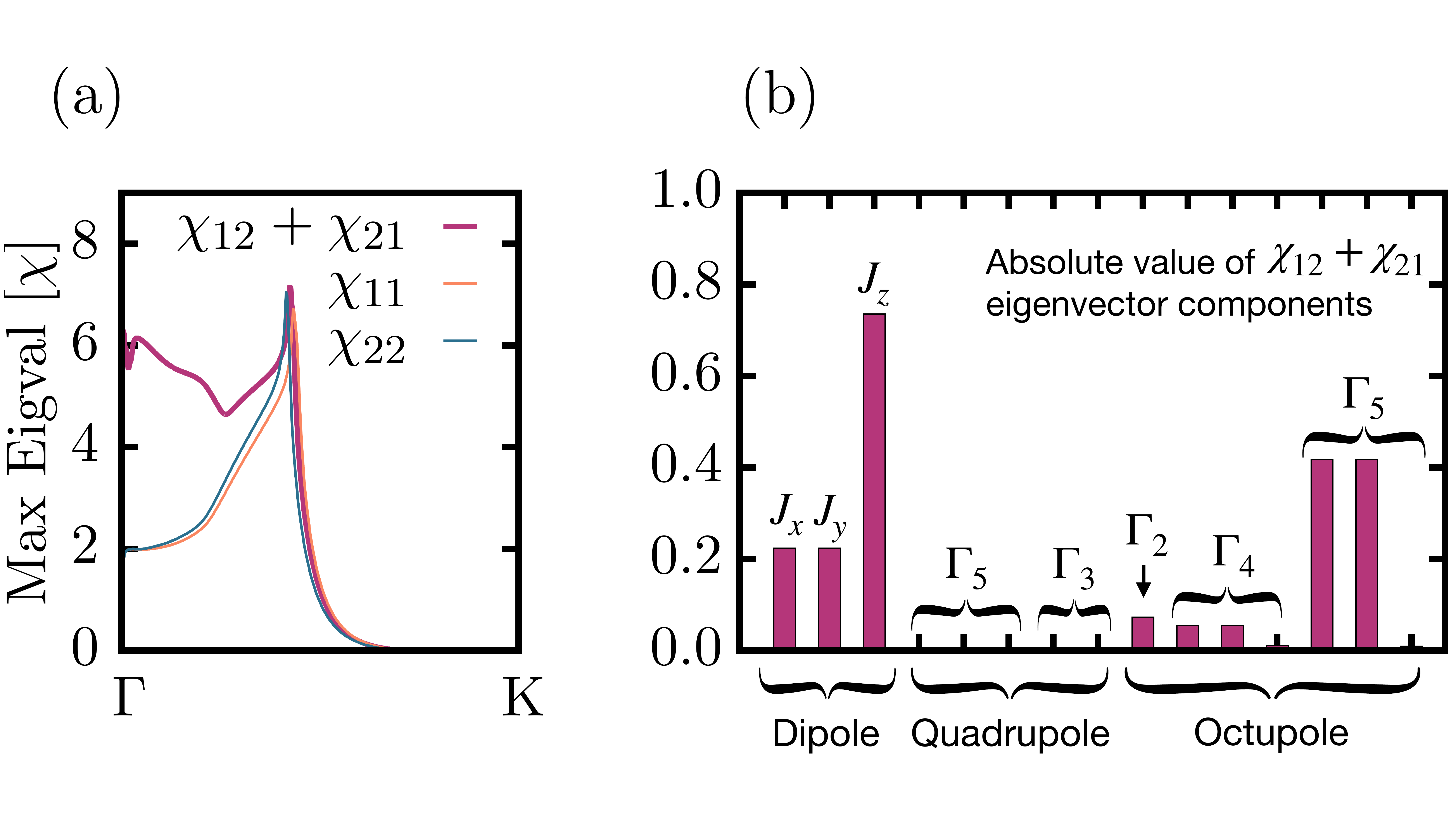}
\vskip -0.5cm
\caption{(a) Maximum eigenvalues of the susceptibility matrix for $j=3/2$ multipoles,
showing that the interband instability (thick solid line) slightly dominates over the intraband ones
(thin solid lines). (b) Magnitude of the eigenvector components for $\chi_{12}\!+\!\chi_{21}$ at $\bq_s$ (peak momentum in (a)), showing
that the dominant order is composed of intertwined time-reversal breaking dipolar ($\sim\!J_z$) and octupolar ($\Gamma_5$) modulations.}
\label{fig5}
\end{figure}

The susceptibility matrix for this $k\!\cdot\! p$ model involves the full set of multipole operators
constructed from the $j\!=\!3/2$ quartet. In addition to
dipole operators $\hJ_i$, this includes
five quadrupoles (triplet: $\Gamma_5$, doublet: $\Gamma_3$) and seven octupoles 
(singlet: $\Gamma_2$, triplets: $\Gamma_4$ and  $\Gamma_5$); see \cite{suppmat} for the
table of operators.
We find that the susceptibility restricted to only the dipole operators reproduces the behavior shown in Fig.~4b
\cite{suppmat}. Fig.~\ref{fig5}a shows the
largest eigenvalue of the {\it full} $15\times 15$ matrix susceptibility,
for both intraband $\chi_{11}, \chi_{22}$ and interband $\chi_{12}+\chi_{21}$
cases, for momentum along the nesting $\Gamma$-$K$ direction.  The strongest
response still occurs for the interband $\chi$ between the two spin-split FSs. The key difference is that the intraband $\chi_{11},\chi_{22}$ 
also show strongly enhanced, but nevertheless subdominant, peaks.
Fig.~\ref{fig5}b plots the magnitude of the eigenvector components of $\chi_{12}+\chi_{21}$
for the different multipoles at the peak momentum, showing that the symmetry breaking corresponds to intertwined magnetic dipolar and octupolar orders. 
As a consequence, the 2DEG stripe should exhibit modulated
loop currents from octupolar order \cite{Paramekanti2020,Voleti2020}.

We conclude with a discussion of the implications of our work.  Given the involvement of EuO, and the fact that a magnetic
stripe model gives a more natural explanation for a single-$\bq$ state than a charge-only model, we conclude that
the `nematic' phase seen by Liu {\it et al.}~\cite{liu} below $T_s \!\sim\! 2.2$\,K is a magnetic stripe.
Experiments such as linear optical dichroism,
Raman scattering, and resonant x-ray diffraction could shed further light on the nature of this broken symmetry state at the interface.
The vestigial nematicity associated with the stripe order is expected to have relatively longer range correlations \cite{Nie2017,Fernandes2019},
so it is likely to be pinned by the device geometry.
The magnetic stripe state
should also strongly impact the SC order.
The gapped FS shown in Fig.~3 will reduce the number of states available for pairing, resulting in a lower $T_c$ in samples
which exhibit nematicity. In addition, it will cause the SC itself to be anisotropic. Furthermore,
for a spin-density wave state, the pairing in a paramagnetic basis can be written as a linear combination of
a spin singlet at $\bQ=0$ (here, $\bQ$ refers to the center of mass momentum of the pair),
and one component of a spin triplet at the magnetic wavevector, $\bQ=\bq_s$ \cite{fenton}.  As such, the 
superconducting state will also exhibit spatial modulation, similar to a pair density wave state \cite{pdw},
as suggested by Liu {\it et al.}~\cite{liu}. A full treatment of the pairing problem
would involve explicitly considering the Rashba splitting, given that its value exceeds both $T_s$ and $T_c$.
Finally, the increase of $T_c$ with carrier density could be due to the suppression of the magnetic stripe phase.
This competition is commonly observed in density wave materials such as Cu$_x$TiSe$_2$ \cite{morosan2006}.

The authors thank Anand Bhattacharya, Changjiang Liu and Ivar Martin for discussions.
This work was supported by the Materials Sciences and Engineering
Division, Basic Energy Sciences, Office of Science, U.~S.~Dept.~of Energy.
AP acknowledges funding from NSERC of Canada.

\bibliography{KTO111}

\begin{thebibliography}{44}%
\makeatletter
\providecommand \@ifxundefined [1]{%
 \@ifx{#1\undefined}
}%
\providecommand \@ifnum [1]{%
 \ifnum #1\expandafter \@firstoftwo
 \else \expandafter \@secondoftwo
 \fi
}%
\providecommand \@ifx [1]{%
 \ifx #1\expandafter \@firstoftwo
 \else \expandafter \@secondoftwo
 \fi
}%
\providecommand \natexlab [1]{#1}%
\providecommand \enquote  [1]{``#1''}%
\providecommand \bibnamefont  [1]{#1}%
\providecommand \bibfnamefont [1]{#1}%
\providecommand \citenamefont [1]{#1}%
\providecommand \href@noop [0]{\@secondoftwo}%
\providecommand \href [0]{\begingroup \@sanitize@url \@href}%
\providecommand \@href[1]{\@@startlink{#1}\@@href}%
\providecommand \@@href[1]{\endgroup#1\@@endlink}%
\providecommand \@sanitize@url [0]{\catcode `\\12\catcode `\$12\catcode
  `\&12\catcode `\#12\catcode `\^12\catcode `\_12\catcode `\%12\relax}%
\providecommand \@@startlink[1]{}%
\providecommand \@@endlink[0]{}%
\providecommand \url  [0]{\begingroup\@sanitize@url \@url }%
\providecommand \@url [1]{\endgroup\@href {#1}{\urlprefix }}%
\providecommand \urlprefix  [0]{URL }%
\providecommand \Eprint [0]{\href }%
\providecommand \doibase [0]{http://dx.doi.org/}%
\providecommand \selectlanguage [0]{\@gobble}%
\providecommand \bibinfo  [0]{\@secondoftwo}%
\providecommand \bibfield  [0]{\@secondoftwo}%
\providecommand \translation [1]{[#1]}%
\providecommand \BibitemOpen [0]{}%
\providecommand \bibitemStop [0]{}%
\providecommand \bibitemNoStop [0]{.\EOS\space}%
\providecommand \EOS [0]{\spacefactor3000\relax}%
\providecommand \BibitemShut  [1]{\csname bibitem#1\endcsname}%
\let\auto@bib@innerbib\@empty
\bibitem [{\citenamefont {Reyren}\ \emph {et~al.}(2007)\citenamefont {Reyren},
  \citenamefont {Thiel}, \citenamefont {Caviglia}, \citenamefont {Kourkoutis},
  \citenamefont {Hammerl}, \citenamefont {Richter}, \citenamefont {Schneider},
  \citenamefont {Kopp}, \citenamefont {R{\"u}etschi}, \citenamefont {Jaccard},
  \citenamefont {Gabay}, \citenamefont {Muller}, \citenamefont {Triscone},\
  and\ \citenamefont {Mannhart}}]{reyren}%
  \BibitemOpen
  \bibfield  {author} {\bibinfo {author} {\bibfnamefont {N.}~\bibnamefont
  {Reyren}}, \bibinfo {author} {\bibfnamefont {S.}~\bibnamefont {Thiel}},
  \bibinfo {author} {\bibfnamefont {A.~D.}\ \bibnamefont {Caviglia}}, \bibinfo
  {author} {\bibfnamefont {L.~F.}\ \bibnamefont {Kourkoutis}}, \bibinfo
  {author} {\bibfnamefont {G.}~\bibnamefont {Hammerl}}, \bibinfo {author}
  {\bibfnamefont {C.}~\bibnamefont {Richter}}, \bibinfo {author} {\bibfnamefont
  {C.~W.}\ \bibnamefont {Schneider}}, \bibinfo {author} {\bibfnamefont
  {T.}~\bibnamefont {Kopp}}, \bibinfo {author} {\bibfnamefont {A.-S.}\
  \bibnamefont {R{\"u}etschi}}, \bibinfo {author} {\bibfnamefont
  {D.}~\bibnamefont {Jaccard}}, \bibinfo {author} {\bibfnamefont
  {M.}~\bibnamefont {Gabay}}, \bibinfo {author} {\bibfnamefont {D.~A.}\
  \bibnamefont {Muller}}, \bibinfo {author} {\bibfnamefont {J.-M.}\
  \bibnamefont {Triscone}}, \ and\ \bibinfo {author} {\bibfnamefont
  {J.}~\bibnamefont {Mannhart}},\ }\href {\doibase 10.1126/science.1146006}
  {\bibfield  {journal} {\bibinfo  {journal} {Science}\ }\textbf {\bibinfo
  {volume} {317}},\ \bibinfo {pages} {1196} (\bibinfo {year}
  {2007})}\BibitemShut {NoStop}%
\bibitem [{\citenamefont {Ueno}\ \emph {et~al.}(2011)\citenamefont {Ueno},
  \citenamefont {Nakamura}, \citenamefont {Shimotani}, \citenamefont {Yuan},
  \citenamefont {Kimura}, \citenamefont {Nojima}, \citenamefont {Aoki},
  \citenamefont {Iwasa},\ and\ \citenamefont {Kawasaki}}]{ueno}%
  \BibitemOpen
  \bibfield  {author} {\bibinfo {author} {\bibfnamefont {K.}~\bibnamefont
  {Ueno}}, \bibinfo {author} {\bibfnamefont {S.}~\bibnamefont {Nakamura}},
  \bibinfo {author} {\bibfnamefont {H.}~\bibnamefont {Shimotani}}, \bibinfo
  {author} {\bibfnamefont {H.}~\bibnamefont {Yuan}}, \bibinfo {author}
  {\bibfnamefont {N.}~\bibnamefont {Kimura}}, \bibinfo {author} {\bibfnamefont
  {T.}~\bibnamefont {Nojima}}, \bibinfo {author} {\bibfnamefont
  {H.}~\bibnamefont {Aoki}}, \bibinfo {author} {\bibfnamefont {Y.}~\bibnamefont
  {Iwasa}}, \ and\ \bibinfo {author} {\bibfnamefont {M.}~\bibnamefont
  {Kawasaki}},\ }\href {https://doi.org/10.1038/nnano.2011.78} {\bibfield
  {journal} {\bibinfo  {journal} {Nature Nanotechnology}\ }\textbf {\bibinfo
  {volume} {6}},\ \bibinfo {pages} {408} (\bibinfo {year} {2011})}\BibitemShut
  {NoStop}%
\bibitem [{\citenamefont {Liu}\ \emph {et~al.}(2020)\citenamefont {Liu},
  \citenamefont {Yan}, \citenamefont {Jin}, \citenamefont {Ma}, \citenamefont
  {Hsiao}, \citenamefont {Lin}, \citenamefont {Bretz-Sullivan}, \citenamefont
  {Zhou}, \citenamefont {Pearson}, \citenamefont {Fisher} \emph
  {et~al.}}]{liu}%
  \BibitemOpen
  \bibfield  {author} {\bibinfo {author} {\bibfnamefont {C.}~\bibnamefont
  {Liu}}, \bibinfo {author} {\bibfnamefont {X.}~\bibnamefont {Yan}}, \bibinfo
  {author} {\bibfnamefont {D.}~\bibnamefont {Jin}}, \bibinfo {author}
  {\bibfnamefont {Y.}~\bibnamefont {Ma}}, \bibinfo {author} {\bibfnamefont
  {H.-W.}\ \bibnamefont {Hsiao}}, \bibinfo {author} {\bibfnamefont
  {Y.}~\bibnamefont {Lin}}, \bibinfo {author} {\bibfnamefont {T.~M.}\
  \bibnamefont {Bretz-Sullivan}}, \bibinfo {author} {\bibfnamefont
  {X.}~\bibnamefont {Zhou}}, \bibinfo {author} {\bibfnamefont {J.}~\bibnamefont
  {Pearson}}, \bibinfo {author} {\bibfnamefont {B.}~\bibnamefont {Fisher}},
  \emph {et~al.},\ }\href {https://arxiv.org/abs/2004.07416} {\bibfield
  {journal} {\bibinfo  {journal} {arXiv preprint arXiv:2004.07416}\ } (\bibinfo
  {year} {2020})}\BibitemShut {NoStop}%
\bibitem [{\citenamefont {Ando}\ \emph {et~al.}(2002)\citenamefont {Ando},
  \citenamefont {Segawa}, \citenamefont {Komiya},\ and\ \citenamefont
  {Lavrov}}]{cuprates_nem_1}%
  \BibitemOpen
  \bibfield  {author} {\bibinfo {author} {\bibfnamefont {Y.}~\bibnamefont
  {Ando}}, \bibinfo {author} {\bibfnamefont {K.}~\bibnamefont {Segawa}},
  \bibinfo {author} {\bibfnamefont {S.}~\bibnamefont {Komiya}}, \ and\ \bibinfo
  {author} {\bibfnamefont {A.~N.}\ \bibnamefont {Lavrov}},\ }\href {\doibase
  10.1103/PhysRevLett.88.137005} {\bibfield  {journal} {\bibinfo  {journal}
  {Phys. Rev. Lett.}\ }\textbf {\bibinfo {volume} {88}},\ \bibinfo {pages}
  {137005} (\bibinfo {year} {2002})}\BibitemShut {NoStop}%
\bibitem [{\citenamefont {Hinkov}\ \emph {et~al.}(2008)\citenamefont {Hinkov},
  \citenamefont {Haug}, \citenamefont {Fauqu{\'e}}, \citenamefont {Bourges},
  \citenamefont {Sidis}, \citenamefont {Ivanov}, \citenamefont {Bernhard},
  \citenamefont {Lin},\ and\ \citenamefont {Keimer}}]{cuprates_nem_2}%
  \BibitemOpen
  \bibfield  {author} {\bibinfo {author} {\bibfnamefont {V.}~\bibnamefont
  {Hinkov}}, \bibinfo {author} {\bibfnamefont {D.}~\bibnamefont {Haug}},
  \bibinfo {author} {\bibfnamefont {B.}~\bibnamefont {Fauqu{\'e}}}, \bibinfo
  {author} {\bibfnamefont {P.}~\bibnamefont {Bourges}}, \bibinfo {author}
  {\bibfnamefont {Y.}~\bibnamefont {Sidis}}, \bibinfo {author} {\bibfnamefont
  {A.}~\bibnamefont {Ivanov}}, \bibinfo {author} {\bibfnamefont
  {C.}~\bibnamefont {Bernhard}}, \bibinfo {author} {\bibfnamefont {C.~T.}\
  \bibnamefont {Lin}}, \ and\ \bibinfo {author} {\bibfnamefont
  {B.}~\bibnamefont {Keimer}},\ }\href {\doibase 10.1126/science.1152309}
  {\bibfield  {journal} {\bibinfo  {journal} {Science}\ }\textbf {\bibinfo
  {volume} {319}},\ \bibinfo {pages} {597} (\bibinfo {year}
  {2008})}\BibitemShut {NoStop}%
\bibitem [{\citenamefont {Paglione}\ and\ \citenamefont
  {Greene}(2010)}]{Paglione_IBSC}%
  \BibitemOpen
  \bibfield  {author} {\bibinfo {author} {\bibfnamefont {J.}~\bibnamefont
  {Paglione}}\ and\ \bibinfo {author} {\bibfnamefont {R.~L.}\ \bibnamefont
  {Greene}},\ }\href {\doibase 10.1038/nphys1759} {\bibfield  {journal}
  {\bibinfo  {journal} {Nature Physics}\ }\textbf {\bibinfo {volume} {6}},\
  \bibinfo {pages} {645} (\bibinfo {year} {2010})}\BibitemShut {NoStop}%
\bibitem [{\citenamefont {Matano}\ \emph {et~al.}(2016)\citenamefont {Matano},
  \citenamefont {Kriener}, \citenamefont {Segawa}, \citenamefont {Ando},\ and\
  \citenamefont {Zheng}}]{bi2se3_1}%
  \BibitemOpen
  \bibfield  {author} {\bibinfo {author} {\bibfnamefont {K.}~\bibnamefont
  {Matano}}, \bibinfo {author} {\bibfnamefont {M.}~\bibnamefont {Kriener}},
  \bibinfo {author} {\bibfnamefont {K.}~\bibnamefont {Segawa}}, \bibinfo
  {author} {\bibfnamefont {Y.}~\bibnamefont {Ando}}, \ and\ \bibinfo {author}
  {\bibfnamefont {G.-q.}\ \bibnamefont {Zheng}},\ }\href {\doibase
  10.1038/nphys3781} {\bibfield  {journal} {\bibinfo  {journal} {Nature
  Physics}\ }\textbf {\bibinfo {volume} {12}},\ \bibinfo {pages} {852}
  (\bibinfo {year} {2016})}\BibitemShut {NoStop}%
\bibitem [{\citenamefont {Yonezawa}\ \emph {et~al.}(2017)\citenamefont
  {Yonezawa}, \citenamefont {Tajiri}, \citenamefont {Nakata}, \citenamefont
  {Nagai}, \citenamefont {Wang}, \citenamefont {Segawa}, \citenamefont {Ando},\
  and\ \citenamefont {Maeno}}]{bi2se3_2}%
  \BibitemOpen
  \bibfield  {author} {\bibinfo {author} {\bibfnamefont {S.}~\bibnamefont
  {Yonezawa}}, \bibinfo {author} {\bibfnamefont {K.}~\bibnamefont {Tajiri}},
  \bibinfo {author} {\bibfnamefont {S.}~\bibnamefont {Nakata}}, \bibinfo
  {author} {\bibfnamefont {Y.}~\bibnamefont {Nagai}}, \bibinfo {author}
  {\bibfnamefont {Z.}~\bibnamefont {Wang}}, \bibinfo {author} {\bibfnamefont
  {K.}~\bibnamefont {Segawa}}, \bibinfo {author} {\bibfnamefont
  {Y.}~\bibnamefont {Ando}}, \ and\ \bibinfo {author} {\bibfnamefont
  {Y.}~\bibnamefont {Maeno}},\ }\href {\doibase 10.1038/nphys3907} {\bibfield
  {journal} {\bibinfo  {journal} {Nature Physics}\ }\textbf {\bibinfo {volume}
  {13}},\ \bibinfo {pages} {123} (\bibinfo {year} {2017})}\BibitemShut
  {NoStop}%
\bibitem [{\citenamefont {Hecker}\ and\ \citenamefont
  {Schmalian}(2018)}]{bi2se3_3}%
  \BibitemOpen
  \bibfield  {author} {\bibinfo {author} {\bibfnamefont {M.}~\bibnamefont
  {Hecker}}\ and\ \bibinfo {author} {\bibfnamefont {J.}~\bibnamefont
  {Schmalian}},\ }\href {\doibase 10.1038/s41535-018-0098-z} {\bibfield
  {journal} {\bibinfo  {journal} {npj Quantum Materials}\ }\textbf {\bibinfo
  {volume} {3}},\ \bibinfo {pages} {1} (\bibinfo {year} {2018})}\BibitemShut
  {NoStop}%
\bibitem [{\citenamefont {Cao}\ \emph {et~al.}(2020)\citenamefont {Cao},
  \citenamefont {Rodan-Legrain}, \citenamefont {Park}, \citenamefont {Yuan},
  \citenamefont {Watanabe}, \citenamefont {Taniguchi}, \citenamefont
  {Fernandes}, \citenamefont {Fu},\ and\ \citenamefont
  {Jarillo-Herrero}}]{TBG_nematic}%
  \BibitemOpen
  \bibfield  {author} {\bibinfo {author} {\bibfnamefont {Y.}~\bibnamefont
  {Cao}}, \bibinfo {author} {\bibfnamefont {D.}~\bibnamefont {Rodan-Legrain}},
  \bibinfo {author} {\bibfnamefont {J.~M.}\ \bibnamefont {Park}}, \bibinfo
  {author} {\bibfnamefont {F.~N.}\ \bibnamefont {Yuan}}, \bibinfo {author}
  {\bibfnamefont {K.}~\bibnamefont {Watanabe}}, \bibinfo {author}
  {\bibfnamefont {T.}~\bibnamefont {Taniguchi}}, \bibinfo {author}
  {\bibfnamefont {R.~M.}\ \bibnamefont {Fernandes}}, \bibinfo {author}
  {\bibfnamefont {L.}~\bibnamefont {Fu}}, \ and\ \bibinfo {author}
  {\bibfnamefont {P.}~\bibnamefont {Jarillo-Herrero}},\ }\href
  {https://arxiv.org/abs/2004.04148} {\bibfield  {journal} {\bibinfo  {journal}
  {arXiv preprint arXiv:2004.04148}\ } (\bibinfo {year} {2020})}\BibitemShut
  {NoStop}%
\bibitem [{\citenamefont {Mattheiss}(1972)}]{Matthias_1972}%
  \BibitemOpen
  \bibfield  {author} {\bibinfo {author} {\bibfnamefont {L.~F.}\ \bibnamefont
  {Mattheiss}},\ }\href {\doibase 10.1103/PhysRevB.6.4718} {\bibfield
  {journal} {\bibinfo  {journal} {Phys. Rev. B}\ }\textbf {\bibinfo {volume}
  {6}},\ \bibinfo {pages} {4718} (\bibinfo {year} {1972})}\BibitemShut
  {NoStop}%
\bibitem [{\citenamefont {Bruno}\ \emph {et~al.}(2019)\citenamefont {Bruno},
  \citenamefont {McKeown~Walker}, \citenamefont {Riccò}, \citenamefont {de~la
  Torre}, \citenamefont {Wang}, \citenamefont {Tamai}, \citenamefont {Kim},
  \citenamefont {Hoesch}, \citenamefont {Bahramy},\ and\ \citenamefont
  {Baumberger}}]{bruno}%
  \BibitemOpen
  \bibfield  {author} {\bibinfo {author} {\bibfnamefont {F.~Y.}\ \bibnamefont
  {Bruno}}, \bibinfo {author} {\bibfnamefont {S.}~\bibnamefont
  {McKeown~Walker}}, \bibinfo {author} {\bibfnamefont {S.}~\bibnamefont
  {Riccò}}, \bibinfo {author} {\bibfnamefont {A.}~\bibnamefont {de~la Torre}},
  \bibinfo {author} {\bibfnamefont {Z.}~\bibnamefont {Wang}}, \bibinfo {author}
  {\bibfnamefont {A.}~\bibnamefont {Tamai}}, \bibinfo {author} {\bibfnamefont
  {T.~K.}\ \bibnamefont {Kim}}, \bibinfo {author} {\bibfnamefont
  {M.}~\bibnamefont {Hoesch}}, \bibinfo {author} {\bibfnamefont {M.~S.}\
  \bibnamefont {Bahramy}}, \ and\ \bibinfo {author} {\bibfnamefont
  {F.}~\bibnamefont {Baumberger}},\ }\href {\doibase 10.1002/aelm.201800860}
  {\bibfield  {journal} {\bibinfo  {journal} {Advanced Electronic Materials}\
  }\textbf {\bibinfo {volume} {5}},\ \bibinfo {pages} {1800860} (\bibinfo
  {year} {2019})}\BibitemShut {NoStop}%
\bibitem [{\citenamefont {Fu}(2015)}]{LiangFu_PRL2015}%
  \BibitemOpen
  \bibfield  {author} {\bibinfo {author} {\bibfnamefont {L.}~\bibnamefont
  {Fu}},\ }\href {\doibase 10.1103/PhysRevLett.115.026401} {\bibfield
  {journal} {\bibinfo  {journal} {Phys. Rev. Lett.}\ }\textbf {\bibinfo
  {volume} {115}},\ \bibinfo {pages} {026401} (\bibinfo {year}
  {2015})}\BibitemShut {NoStop}%
\bibitem [{\citenamefont {Harter}\ \emph {et~al.}(2017)\citenamefont {Harter},
  \citenamefont {Zhao}, \citenamefont {Yan}, \citenamefont {Mandrus},\ and\
  \citenamefont {Hsieh}}]{Harter295}%
  \BibitemOpen
  \bibfield  {author} {\bibinfo {author} {\bibfnamefont {J.~W.}\ \bibnamefont
  {Harter}}, \bibinfo {author} {\bibfnamefont {Z.~Y.}\ \bibnamefont {Zhao}},
  \bibinfo {author} {\bibfnamefont {J.-Q.}\ \bibnamefont {Yan}}, \bibinfo
  {author} {\bibfnamefont {D.~G.}\ \bibnamefont {Mandrus}}, \ and\ \bibinfo
  {author} {\bibfnamefont {D.}~\bibnamefont {Hsieh}},\ }\href {\doibase
  10.1126/science.aad1188} {\bibfield  {journal} {\bibinfo  {journal}
  {Science}\ }\textbf {\bibinfo {volume} {356}},\ \bibinfo {pages} {295}
  (\bibinfo {year} {2017})}\BibitemShut {NoStop}%
\bibitem [{\citenamefont {Di~Matteo}\ and\ \citenamefont
  {Norman}(2017)}]{Norman_PRB2017}%
  \BibitemOpen
  \bibfield  {author} {\bibinfo {author} {\bibfnamefont {S.}~\bibnamefont
  {Di~Matteo}}\ and\ \bibinfo {author} {\bibfnamefont {M.~R.}\ \bibnamefont
  {Norman}},\ }\href {\doibase 10.1103/PhysRevB.96.115156} {\bibfield
  {journal} {\bibinfo  {journal} {Phys. Rev. B}\ }\textbf {\bibinfo {volume}
  {96}},\ \bibinfo {pages} {115156} (\bibinfo {year} {2017})}\BibitemShut
  {NoStop}%
\bibitem [{\citenamefont {Hayami}\ \emph {et~al.}(2019)\citenamefont {Hayami},
  \citenamefont {Yanagi}, \citenamefont {Kusunose},\ and\ \citenamefont
  {Motome}}]{hayami2019}%
  \BibitemOpen
  \bibfield  {author} {\bibinfo {author} {\bibfnamefont {S.}~\bibnamefont
  {Hayami}}, \bibinfo {author} {\bibfnamefont {Y.}~\bibnamefont {Yanagi}},
  \bibinfo {author} {\bibfnamefont {H.}~\bibnamefont {Kusunose}}, \ and\
  \bibinfo {author} {\bibfnamefont {Y.}~\bibnamefont {Motome}},\ }\href
  {\doibase 10.1103/PhysRevLett.122.147602} {\bibfield  {journal} {\bibinfo
  {journal} {Phys. Rev. Lett.}\ }\textbf {\bibinfo {volume} {122}},\ \bibinfo
  {pages} {147602} (\bibinfo {year} {2019})}\BibitemShut {NoStop}%
\bibitem [{\citenamefont {Jellison}\ \emph {et~al.}(2006)\citenamefont
  {Jellison}, \citenamefont {Paulauskas}, \citenamefont {Boatner},\ and\
  \citenamefont {Singh}}]{Jellison_PRB2006}%
  \BibitemOpen
  \bibfield  {author} {\bibinfo {author} {\bibfnamefont {G.~E.}\ \bibnamefont
  {Jellison}}, \bibinfo {author} {\bibfnamefont {I.}~\bibnamefont
  {Paulauskas}}, \bibinfo {author} {\bibfnamefont {L.~A.}\ \bibnamefont
  {Boatner}}, \ and\ \bibinfo {author} {\bibfnamefont {D.~J.}\ \bibnamefont
  {Singh}},\ }\href {\doibase 10.1103/PhysRevB.74.155130} {\bibfield  {journal}
  {\bibinfo  {journal} {Phys. Rev. B}\ }\textbf {\bibinfo {volume} {74}},\
  \bibinfo {pages} {155130} (\bibinfo {year} {2006})}\BibitemShut {NoStop}%
\bibitem [{\citenamefont {Nakamura}\ and\ \citenamefont
  {Kimura}(2009)}]{Nakamura_PRB2009}%
  \BibitemOpen
  \bibfield  {author} {\bibinfo {author} {\bibfnamefont {H.}~\bibnamefont
  {Nakamura}}\ and\ \bibinfo {author} {\bibfnamefont {T.}~\bibnamefont
  {Kimura}},\ }\href {\doibase 10.1103/PhysRevB.80.121308} {\bibfield
  {journal} {\bibinfo  {journal} {Phys. Rev. B}\ }\textbf {\bibinfo {volume}
  {80}},\ \bibinfo {pages} {121308} (\bibinfo {year} {2009})}\BibitemShut
  {NoStop}%
\bibitem [{\citenamefont {King}\ \emph {et~al.}(2012)\citenamefont {King},
  \citenamefont {He}, \citenamefont {Eknapakul}, \citenamefont {Buaphet},
  \citenamefont {Mo}, \citenamefont {Kaneko}, \citenamefont {Harashima},
  \citenamefont {Hikita}, \citenamefont {Bahramy}, \citenamefont {Bell},
  \citenamefont {Hussain}, \citenamefont {Tokura}, \citenamefont {Shen},
  \citenamefont {Hwang}, \citenamefont {Baumberger},\ and\ \citenamefont
  {Meevasana}}]{King_PRL2012}%
  \BibitemOpen
  \bibfield  {author} {\bibinfo {author} {\bibfnamefont {P.~D.~C.}\
  \bibnamefont {King}}, \bibinfo {author} {\bibfnamefont {R.~H.}\ \bibnamefont
  {He}}, \bibinfo {author} {\bibfnamefont {T.}~\bibnamefont {Eknapakul}},
  \bibinfo {author} {\bibfnamefont {P.}~\bibnamefont {Buaphet}}, \bibinfo
  {author} {\bibfnamefont {S.-K.}\ \bibnamefont {Mo}}, \bibinfo {author}
  {\bibfnamefont {Y.}~\bibnamefont {Kaneko}}, \bibinfo {author} {\bibfnamefont
  {S.}~\bibnamefont {Harashima}}, \bibinfo {author} {\bibfnamefont
  {Y.}~\bibnamefont {Hikita}}, \bibinfo {author} {\bibfnamefont {M.~S.}\
  \bibnamefont {Bahramy}}, \bibinfo {author} {\bibfnamefont {C.}~\bibnamefont
  {Bell}}, \bibinfo {author} {\bibfnamefont {Z.}~\bibnamefont {Hussain}},
  \bibinfo {author} {\bibfnamefont {Y.}~\bibnamefont {Tokura}}, \bibinfo
  {author} {\bibfnamefont {Z.-X.}\ \bibnamefont {Shen}}, \bibinfo {author}
  {\bibfnamefont {H.~Y.}\ \bibnamefont {Hwang}}, \bibinfo {author}
  {\bibfnamefont {F.}~\bibnamefont {Baumberger}}, \ and\ \bibinfo {author}
  {\bibfnamefont {W.}~\bibnamefont {Meevasana}},\ }\href {\doibase
  10.1103/PhysRevLett.108.117602} {\bibfield  {journal} {\bibinfo  {journal}
  {Phys. Rev. Lett.}\ }\textbf {\bibinfo {volume} {108}},\ \bibinfo {pages}
  {117602} (\bibinfo {year} {2012})}\BibitemShut {NoStop}%
\bibitem [{\citenamefont {Santander-Syro}\ \emph {et~al.}(2012)\citenamefont
  {Santander-Syro}, \citenamefont {Bareille}, \citenamefont {Fortuna},
  \citenamefont {Copie}, \citenamefont {Gabay}, \citenamefont {Bertran},
  \citenamefont {Taleb-Ibrahimi}, \citenamefont {Le~F\`evre}, \citenamefont
  {Herranz}, \citenamefont {Reyren}, \citenamefont {Bibes}, \citenamefont
  {Barth\'el\'emy}, \citenamefont {Lecoeur}, \citenamefont {Guevara},\ and\
  \citenamefont {Rozenberg}}]{Rozenberg_PRB2012}%
  \BibitemOpen
  \bibfield  {author} {\bibinfo {author} {\bibfnamefont {A.~F.}\ \bibnamefont
  {Santander-Syro}}, \bibinfo {author} {\bibfnamefont {C.}~\bibnamefont
  {Bareille}}, \bibinfo {author} {\bibfnamefont {F.}~\bibnamefont {Fortuna}},
  \bibinfo {author} {\bibfnamefont {O.}~\bibnamefont {Copie}}, \bibinfo
  {author} {\bibfnamefont {M.}~\bibnamefont {Gabay}}, \bibinfo {author}
  {\bibfnamefont {F.}~\bibnamefont {Bertran}}, \bibinfo {author} {\bibfnamefont
  {A.}~\bibnamefont {Taleb-Ibrahimi}}, \bibinfo {author} {\bibfnamefont
  {P.}~\bibnamefont {Le~F\`evre}}, \bibinfo {author} {\bibfnamefont
  {G.}~\bibnamefont {Herranz}}, \bibinfo {author} {\bibfnamefont
  {N.}~\bibnamefont {Reyren}}, \bibinfo {author} {\bibfnamefont
  {M.}~\bibnamefont {Bibes}}, \bibinfo {author} {\bibfnamefont
  {A.}~\bibnamefont {Barth\'el\'emy}}, \bibinfo {author} {\bibfnamefont
  {P.}~\bibnamefont {Lecoeur}}, \bibinfo {author} {\bibfnamefont
  {J.}~\bibnamefont {Guevara}}, \ and\ \bibinfo {author} {\bibfnamefont
  {M.~J.}\ \bibnamefont {Rozenberg}},\ }\href {\doibase
  10.1103/PhysRevB.86.121107} {\bibfield  {journal} {\bibinfo  {journal} {Phys.
  Rev. B}\ }\textbf {\bibinfo {volume} {86}},\ \bibinfo {pages} {121107}
  (\bibinfo {year} {2012})}\BibitemShut {NoStop}%
\bibitem [{\citenamefont {Zhang}\ \emph {et~al.}(2019)\citenamefont {Zhang},
  \citenamefont {Yan}, \citenamefont {Zhang}, \citenamefont {Wang},
  \citenamefont {Xiong}, \citenamefont {Zhang}, \citenamefont {Qi},
  \citenamefont {Zhang}, \citenamefont {Han}, \citenamefont {Wu}, \citenamefont
  {Liu}, \citenamefont {Chen}, \citenamefont {Shen},\ and\ \citenamefont
  {Sun}}]{Sung_ACSNano2019}%
  \BibitemOpen
  \bibfield  {author} {\bibinfo {author} {\bibfnamefont {H.}~\bibnamefont
  {Zhang}}, \bibinfo {author} {\bibfnamefont {X.}~\bibnamefont {Yan}}, \bibinfo
  {author} {\bibfnamefont {X.}~\bibnamefont {Zhang}}, \bibinfo {author}
  {\bibfnamefont {S.}~\bibnamefont {Wang}}, \bibinfo {author} {\bibfnamefont
  {C.}~\bibnamefont {Xiong}}, \bibinfo {author} {\bibfnamefont
  {H.}~\bibnamefont {Zhang}}, \bibinfo {author} {\bibfnamefont
  {S.}~\bibnamefont {Qi}}, \bibinfo {author} {\bibfnamefont {J.}~\bibnamefont
  {Zhang}}, \bibinfo {author} {\bibfnamefont {F.}~\bibnamefont {Han}}, \bibinfo
  {author} {\bibfnamefont {N.}~\bibnamefont {Wu}}, \bibinfo {author}
  {\bibfnamefont {B.}~\bibnamefont {Liu}}, \bibinfo {author} {\bibfnamefont
  {Y.}~\bibnamefont {Chen}}, \bibinfo {author} {\bibfnamefont {B.}~\bibnamefont
  {Shen}}, \ and\ \bibinfo {author} {\bibfnamefont {J.}~\bibnamefont {Sun}},\
  }\bibfield  {booktitle} {\emph {\bibinfo {booktitle} {ACS Nano}},\ }\href
  {\doibase 10.1021/acsnano.8b07622} {\bibfield  {journal} {\bibinfo  {journal}
  {ACS Nano}\ }\textbf {\bibinfo {volume} {13}},\ \bibinfo {pages} {609}
  (\bibinfo {year} {2019})}\BibitemShut {NoStop}%
\bibitem [{\citenamefont {Bareille}\ \emph {et~al.}(2014)\citenamefont
  {Bareille}, \citenamefont {Fortuna}, \citenamefont {R{\"o}del}, \citenamefont
  {Bertran}, \citenamefont {Gabay}, \citenamefont {Cubelos}, \citenamefont
  {Taleb-Ibrahimi}, \citenamefont {Le~Fevre}, \citenamefont {Bibes},
  \citenamefont {Barth{\'e}l{\'e}my} \emph {et~al.}}]{bareille}%
  \BibitemOpen
  \bibfield  {author} {\bibinfo {author} {\bibfnamefont {C.}~\bibnamefont
  {Bareille}}, \bibinfo {author} {\bibfnamefont {F.}~\bibnamefont {Fortuna}},
  \bibinfo {author} {\bibfnamefont {T.}~\bibnamefont {R{\"o}del}}, \bibinfo
  {author} {\bibfnamefont {F.}~\bibnamefont {Bertran}}, \bibinfo {author}
  {\bibfnamefont {M.}~\bibnamefont {Gabay}}, \bibinfo {author} {\bibfnamefont
  {O.~H.}\ \bibnamefont {Cubelos}}, \bibinfo {author} {\bibfnamefont
  {A.}~\bibnamefont {Taleb-Ibrahimi}}, \bibinfo {author} {\bibfnamefont
  {P.}~\bibnamefont {Le~Fevre}}, \bibinfo {author} {\bibfnamefont
  {M.}~\bibnamefont {Bibes}}, \bibinfo {author} {\bibfnamefont
  {A.}~\bibnamefont {Barth{\'e}l{\'e}my}},  \emph {et~al.},\ }\href {\doibase
  10.1038/srep03586} {\bibfield  {journal} {\bibinfo  {journal} {Scientific
  Reports}\ }\textbf {\bibinfo {volume} {4}},\ \bibinfo {pages} {3586}
  (\bibinfo {year} {2014})}\BibitemShut {NoStop}%
\bibitem [{\citenamefont {Xiao}\ \emph {et~al.}(2011)\citenamefont {Xiao},
  \citenamefont {Zhu}, \citenamefont {Ran}, \citenamefont {Nagaosa},\ and\
  \citenamefont {Okamoto}}]{xiao}%
  \BibitemOpen
  \bibfield  {author} {\bibinfo {author} {\bibfnamefont {D.}~\bibnamefont
  {Xiao}}, \bibinfo {author} {\bibfnamefont {W.}~\bibnamefont {Zhu}}, \bibinfo
  {author} {\bibfnamefont {Y.}~\bibnamefont {Ran}}, \bibinfo {author}
  {\bibfnamefont {N.}~\bibnamefont {Nagaosa}}, \ and\ \bibinfo {author}
  {\bibfnamefont {S.}~\bibnamefont {Okamoto}},\ }\href {\doibase
  10.1038/ncomms1602} {\bibfield  {journal} {\bibinfo  {journal} {Nature
  Communications}\ }\textbf {\bibinfo {volume} {2}},\ \bibinfo {pages} {1}
  (\bibinfo {year} {2011})}\BibitemShut {NoStop}%
\bibitem [{sup()}]{suppmat}%
  \BibitemOpen
  \href@noop {} {}\bibinfo {note} {See the Supplemental Material for details of
  the (i) bilayer tight-binding model, (ii) symmetry analysis leading to
  $k\cdot p$ model and its Fermi surfaces, table of multipolar operators, and
  additional susceptibility plots.}\BibitemShut {Stop}%
\bibitem [{\citenamefont {Khalsa}\ \emph {et~al.}(2013)\citenamefont {Khalsa},
  \citenamefont {Lee},\ and\ \citenamefont {MacDonald}}]{khalsa}%
  \BibitemOpen
  \bibfield  {author} {\bibinfo {author} {\bibfnamefont {G.}~\bibnamefont
  {Khalsa}}, \bibinfo {author} {\bibfnamefont {B.}~\bibnamefont {Lee}}, \ and\
  \bibinfo {author} {\bibfnamefont {A.~H.}\ \bibnamefont {MacDonald}},\ }\href
  {\doibase 10.1103/PhysRevB.88.041302} {\bibfield  {journal} {\bibinfo
  {journal} {Phys. Rev. B}\ }\textbf {\bibinfo {volume} {88}},\ \bibinfo
  {pages} {041302} (\bibinfo {year} {2013})}\BibitemShut {NoStop}%
\bibitem [{1su()}]{1supp}%
  \BibitemOpen
  \href@noop {} {}\bibinfo {note} {These two bands refer to the lowest subset
  of bands in quantum well language. There is weak evidence from ARPES for
  higher level sub-bands, but as these are over $100$ meV higher in energy,
  they play no role in regards to the low energy physics discussed
  here.}\BibitemShut {Stop}%
\bibitem [{\citenamefont {Desclaux}(1973)}]{desclaux}%
  \BibitemOpen
  \bibfield  {author} {\bibinfo {author} {\bibfnamefont {J.}~\bibnamefont
  {Desclaux}},\ }\href@noop {} {\bibfield  {journal} {\bibinfo  {journal}
  {Atomic data and nuclear data tables}\ }\textbf {\bibinfo {volume} {12}},\
  \bibinfo {pages} {311} (\bibinfo {year} {1973})}\BibitemShut {NoStop}%
\bibitem [{\citenamefont {Zhang}\ \emph {et~al.}(2018)\citenamefont {Zhang},
  \citenamefont {Yun}, \citenamefont {Zhang}, \citenamefont {Zhang},
  \citenamefont {Ma}, \citenamefont {Yan}, \citenamefont {Wang}, \citenamefont
  {Li}, \citenamefont {Li}, \citenamefont {Khan}, \citenamefont {Chen},
  \citenamefont {Liu}, \citenamefont {Hu}, \citenamefont {Liu}, \citenamefont
  {Shen}, \citenamefont {Han},\ and\ \citenamefont {Sun}}]{zhang}%
  \BibitemOpen
  \bibfield  {author} {\bibinfo {author} {\bibfnamefont {H.}~\bibnamefont
  {Zhang}}, \bibinfo {author} {\bibfnamefont {Y.}~\bibnamefont {Yun}}, \bibinfo
  {author} {\bibfnamefont {X.}~\bibnamefont {Zhang}}, \bibinfo {author}
  {\bibfnamefont {H.}~\bibnamefont {Zhang}}, \bibinfo {author} {\bibfnamefont
  {Y.}~\bibnamefont {Ma}}, \bibinfo {author} {\bibfnamefont {X.}~\bibnamefont
  {Yan}}, \bibinfo {author} {\bibfnamefont {F.}~\bibnamefont {Wang}}, \bibinfo
  {author} {\bibfnamefont {G.}~\bibnamefont {Li}}, \bibinfo {author}
  {\bibfnamefont {R.}~\bibnamefont {Li}}, \bibinfo {author} {\bibfnamefont
  {T.}~\bibnamefont {Khan}}, \bibinfo {author} {\bibfnamefont {Y.}~\bibnamefont
  {Chen}}, \bibinfo {author} {\bibfnamefont {W.}~\bibnamefont {Liu}}, \bibinfo
  {author} {\bibfnamefont {F.}~\bibnamefont {Hu}}, \bibinfo {author}
  {\bibfnamefont {B.}~\bibnamefont {Liu}}, \bibinfo {author} {\bibfnamefont
  {B.}~\bibnamefont {Shen}}, \bibinfo {author} {\bibfnamefont {W.}~\bibnamefont
  {Han}}, \ and\ \bibinfo {author} {\bibfnamefont {J.}~\bibnamefont {Sun}},\
  }\href {\doibase 10.1103/PhysRevLett.121.116803} {\bibfield  {journal}
  {\bibinfo  {journal} {Phys. Rev. Lett.}\ }\textbf {\bibinfo {volume} {121}},\
  \bibinfo {pages} {116803} (\bibinfo {year} {2018})}\BibitemShut {NoStop}%
\bibitem [{2su()}]{2supp}%
  \BibitemOpen
  \href@noop {} {}\bibinfo {note} {The Zeeman splittings of $\epsilon_{k,n}$
  were not included in the calculations. Also note that in Figs.~2c and 4b, we
  plot the diagonal elements of the 3 by 3 spin matrix. We find this diagonal
  approximation is a good one for the maximum (i.e., $zz$) eigenvalue.
  Moreover, in Fig.~2c, an implicit sum has been done over the Kramers
  degeneracy. That is 11 stands for 11+12+21+22, etc.}\BibitemShut {Stop}%
\bibitem [{\citenamefont {Norman}\ \emph {et~al.}(2007)\citenamefont {Norman},
  \citenamefont {Kanigel}, \citenamefont {Randeria}, \citenamefont
  {Chatterjee},\ and\ \citenamefont {Campuzano}}]{arc}%
  \BibitemOpen
  \bibfield  {author} {\bibinfo {author} {\bibfnamefont {M.~R.}\ \bibnamefont
  {Norman}}, \bibinfo {author} {\bibfnamefont {A.}~\bibnamefont {Kanigel}},
  \bibinfo {author} {\bibfnamefont {M.}~\bibnamefont {Randeria}}, \bibinfo
  {author} {\bibfnamefont {U.}~\bibnamefont {Chatterjee}}, \ and\ \bibinfo
  {author} {\bibfnamefont {J.~C.}\ \bibnamefont {Campuzano}},\ }\href {\doibase
  10.1103/PhysRevB.76.174501} {\bibfield  {journal} {\bibinfo  {journal} {Phys.
  Rev. B}\ }\textbf {\bibinfo {volume} {76}},\ \bibinfo {pages} {174501}
  (\bibinfo {year} {2007})}\BibitemShut {NoStop}%
\bibitem [{\citenamefont {McMillan}(1975)}]{mcmillan}%
  \BibitemOpen
  \bibfield  {author} {\bibinfo {author} {\bibfnamefont {W.~L.}\ \bibnamefont
  {McMillan}},\ }\href {\doibase 10.1103/PhysRevB.12.1187} {\bibfield
  {journal} {\bibinfo  {journal} {Phys. Rev. B}\ }\textbf {\bibinfo {volume}
  {12}},\ \bibinfo {pages} {1187} (\bibinfo {year} {1975})}\BibitemShut
  {NoStop}%
\bibitem [{\citenamefont {Melikyan}\ and\ \citenamefont
  {Norman}(2014)}]{ashot}%
  \BibitemOpen
  \bibfield  {author} {\bibinfo {author} {\bibfnamefont {A.}~\bibnamefont
  {Melikyan}}\ and\ \bibinfo {author} {\bibfnamefont {M.~R.}\ \bibnamefont
  {Norman}},\ }\href {\doibase 10.1103/PhysRevB.89.024507} {\bibfield
  {journal} {\bibinfo  {journal} {Phys. Rev. B}\ }\textbf {\bibinfo {volume}
  {89}},\ \bibinfo {pages} {024507} (\bibinfo {year} {2014})}\BibitemShut
  {NoStop}%
\bibitem [{\citenamefont {Boudjada}\ \emph {et~al.}(2018)\citenamefont
  {Boudjada}, \citenamefont {Wachtel},\ and\ \citenamefont
  {Paramekanti}}]{boudjada}%
  \BibitemOpen
  \bibfield  {author} {\bibinfo {author} {\bibfnamefont {N.}~\bibnamefont
  {Boudjada}}, \bibinfo {author} {\bibfnamefont {G.}~\bibnamefont {Wachtel}}, \
  and\ \bibinfo {author} {\bibfnamefont {A.}~\bibnamefont {Paramekanti}},\
  }\href {\doibase 10.1103/PhysRevLett.120.086802} {\bibfield  {journal}
  {\bibinfo  {journal} {Phys. Rev. Lett.}\ }\textbf {\bibinfo {volume} {120}},\
  \bibinfo {pages} {086802} (\bibinfo {year} {2018})}\BibitemShut {NoStop}%
\bibitem [{\citenamefont {Luttinger}(1956)}]{Luttinger_1956}%
  \BibitemOpen
  \bibfield  {author} {\bibinfo {author} {\bibfnamefont {J.~M.}\ \bibnamefont
  {Luttinger}},\ }\href {\doibase 10.1103/PhysRev.102.1030} {\bibfield
  {journal} {\bibinfo  {journal} {Phys. Rev.}\ }\textbf {\bibinfo {volume}
  {102}},\ \bibinfo {pages} {1030} (\bibinfo {year} {1956})}\BibitemShut
  {NoStop}%
\bibitem [{\citenamefont {Venderbos}\ \emph {et~al.}(2018)\citenamefont
  {Venderbos}, \citenamefont {Savary}, \citenamefont {Ruhman}, \citenamefont
  {Lee},\ and\ \citenamefont {Fu}}]{LiangFu_PRX2018}%
  \BibitemOpen
  \bibfield  {author} {\bibinfo {author} {\bibfnamefont {J.~W.~F.}\
  \bibnamefont {Venderbos}}, \bibinfo {author} {\bibfnamefont {L.}~\bibnamefont
  {Savary}}, \bibinfo {author} {\bibfnamefont {J.}~\bibnamefont {Ruhman}},
  \bibinfo {author} {\bibfnamefont {P.~A.}\ \bibnamefont {Lee}}, \ and\
  \bibinfo {author} {\bibfnamefont {L.}~\bibnamefont {Fu}},\ }\href {\doibase
  10.1103/PhysRevX.8.011029} {\bibfield  {journal} {\bibinfo  {journal} {Phys.
  Rev. X}\ }\textbf {\bibinfo {volume} {8}},\ \bibinfo {pages} {011029}
  (\bibinfo {year} {2018})}\BibitemShut {NoStop}%
\bibitem [{\citenamefont {Paramekanti}\ \emph {et~al.}(2020)\citenamefont
  {Paramekanti}, \citenamefont {Maharaj},\ and\ \citenamefont
  {Gaulin}}]{Paramekanti2020}%
  \BibitemOpen
  \bibfield  {author} {\bibinfo {author} {\bibfnamefont {A.}~\bibnamefont
  {Paramekanti}}, \bibinfo {author} {\bibfnamefont {D.~D.}\ \bibnamefont
  {Maharaj}}, \ and\ \bibinfo {author} {\bibfnamefont {B.~D.}\ \bibnamefont
  {Gaulin}},\ }\href {\doibase 10.1103/PhysRevB.101.054439} {\bibfield
  {journal} {\bibinfo  {journal} {Phys. Rev. B}\ }\textbf {\bibinfo {volume}
  {101}},\ \bibinfo {pages} {054439} (\bibinfo {year} {2020})}\BibitemShut
  {NoStop}%
\bibitem [{\citenamefont {Voleti}\ \emph {et~al.}(2020)\citenamefont {Voleti},
  \citenamefont {Maharaj}, \citenamefont {Gaulin}, \citenamefont {Luke},\ and\
  \citenamefont {Paramekanti}}]{Voleti2020}%
  \BibitemOpen
  \bibfield  {author} {\bibinfo {author} {\bibfnamefont {S.}~\bibnamefont
  {Voleti}}, \bibinfo {author} {\bibfnamefont {D.~D.}\ \bibnamefont {Maharaj}},
  \bibinfo {author} {\bibfnamefont {B.~D.}\ \bibnamefont {Gaulin}}, \bibinfo
  {author} {\bibfnamefont {G.}~\bibnamefont {Luke}}, \ and\ \bibinfo {author}
  {\bibfnamefont {A.}~\bibnamefont {Paramekanti}},\ }\href {\doibase
  10.1103/PhysRevB.101.155118} {\bibfield  {journal} {\bibinfo  {journal}
  {Phys. Rev. B}\ }\textbf {\bibinfo {volume} {101}},\ \bibinfo {pages}
  {155118} (\bibinfo {year} {2020})}\BibitemShut {NoStop}%
\bibitem [{\citenamefont {Nie}\ \emph {et~al.}(2017)\citenamefont {Nie},
  \citenamefont {Maharaj}, \citenamefont {Fradkin},\ and\ \citenamefont
  {Kivelson}}]{Nie2017}%
  \BibitemOpen
  \bibfield  {author} {\bibinfo {author} {\bibfnamefont {L.}~\bibnamefont
  {Nie}}, \bibinfo {author} {\bibfnamefont {A.~V.}\ \bibnamefont {Maharaj}},
  \bibinfo {author} {\bibfnamefont {E.}~\bibnamefont {Fradkin}}, \ and\
  \bibinfo {author} {\bibfnamefont {S.~A.}\ \bibnamefont {Kivelson}},\ }\href
  {\doibase 10.1103/PhysRevB.96.085142} {\bibfield  {journal} {\bibinfo
  {journal} {Phys. Rev. B}\ }\textbf {\bibinfo {volume} {96}},\ \bibinfo
  {pages} {085142} (\bibinfo {year} {2017})}\BibitemShut {NoStop}%
\bibitem [{\citenamefont {Fernandes}\ \emph {et~al.}(2019)\citenamefont
  {Fernandes}, \citenamefont {Orth},\ and\ \citenamefont
  {Schmalian}}]{Fernandes2019}%
  \BibitemOpen
  \bibfield  {author} {\bibinfo {author} {\bibfnamefont {R.~M.}\ \bibnamefont
  {Fernandes}}, \bibinfo {author} {\bibfnamefont {P.~P.}\ \bibnamefont {Orth}},
  \ and\ \bibinfo {author} {\bibfnamefont {J.}~\bibnamefont {Schmalian}},\
  }\href {\doibase 10.1146/annurev-conmatphys-031218-013200} {\bibfield
  {journal} {\bibinfo  {journal} {Annual Review of Condensed Matter Physics}\
  }\textbf {\bibinfo {volume} {10}},\ \bibinfo {pages} {133} (\bibinfo {year}
  {2019})},\ \Eprint
  {http://arxiv.org/abs/https://doi.org/10.1146/annurev-conmatphys-031218-013200}
  {https://doi.org/10.1146/annurev-conmatphys-031218-013200} \BibitemShut
  {NoStop}%
\bibitem [{\citenamefont {Fenton}(1984)}]{fenton}%
  \BibitemOpen
  \bibfield  {author} {\bibinfo {author} {\bibfnamefont {E.~W.}\ \bibnamefont
  {Fenton}},\ }\href {\doibase 10.1143/PTPS.80.94} {\bibfield  {journal}
  {\bibinfo  {journal} {Progress of Theoretical Physics Supplement}\ }\textbf
  {\bibinfo {volume} {80}},\ \bibinfo {pages} {94} (\bibinfo {year}
  {1984})}\BibitemShut {NoStop}%
\bibitem [{\citenamefont {Agterberg}\ \emph {et~al.}(2020)\citenamefont
  {Agterberg}, \citenamefont {Davis}, \citenamefont {Edkins}, \citenamefont
  {Fradkin}, \citenamefont {Van~Harlingen}, \citenamefont {Kivelson},
  \citenamefont {Lee}, \citenamefont {Radzihovsky}, \citenamefont {Tranquada},\
  and\ \citenamefont {Wang}}]{pdw}%
  \BibitemOpen
  \bibfield  {author} {\bibinfo {author} {\bibfnamefont {D.~F.}\ \bibnamefont
  {Agterberg}}, \bibinfo {author} {\bibfnamefont {J.~S.}\ \bibnamefont
  {Davis}}, \bibinfo {author} {\bibfnamefont {S.~D.}\ \bibnamefont {Edkins}},
  \bibinfo {author} {\bibfnamefont {E.}~\bibnamefont {Fradkin}}, \bibinfo
  {author} {\bibfnamefont {D.~J.}\ \bibnamefont {Van~Harlingen}}, \bibinfo
  {author} {\bibfnamefont {S.~A.}\ \bibnamefont {Kivelson}}, \bibinfo {author}
  {\bibfnamefont {P.~A.}\ \bibnamefont {Lee}}, \bibinfo {author} {\bibfnamefont
  {L.}~\bibnamefont {Radzihovsky}}, \bibinfo {author} {\bibfnamefont {J.~M.}\
  \bibnamefont {Tranquada}}, \ and\ \bibinfo {author} {\bibfnamefont
  {Y.}~\bibnamefont {Wang}},\ }\href {\doibase
  10.1146/annurev-conmatphys-031119-050711} {\bibfield  {journal} {\bibinfo
  {journal} {Annual Review of Condensed Matter Physics}\ }\textbf {\bibinfo
  {volume} {11}},\ \bibinfo {pages} {231} (\bibinfo {year} {2020})}\BibitemShut
  {NoStop}%
\bibitem [{\citenamefont {Morosan}\ \emph {et~al.}(2006)\citenamefont
  {Morosan}, \citenamefont {Zandbergen}, \citenamefont {Dennis}, \citenamefont
  {Bos}, \citenamefont {Onose}, \citenamefont {Klimczuk}, \citenamefont
  {Ramirez}, \citenamefont {Ong},\ and\ \citenamefont {Cava}}]{morosan2006}%
  \BibitemOpen
  \bibfield  {author} {\bibinfo {author} {\bibfnamefont {E.}~\bibnamefont
  {Morosan}}, \bibinfo {author} {\bibfnamefont {H.~W.}\ \bibnamefont
  {Zandbergen}}, \bibinfo {author} {\bibfnamefont {B.~S.}\ \bibnamefont
  {Dennis}}, \bibinfo {author} {\bibfnamefont {J.~W.~G.}\ \bibnamefont {Bos}},
  \bibinfo {author} {\bibfnamefont {Y.}~\bibnamefont {Onose}}, \bibinfo
  {author} {\bibfnamefont {T.}~\bibnamefont {Klimczuk}}, \bibinfo {author}
  {\bibfnamefont {A.~P.}\ \bibnamefont {Ramirez}}, \bibinfo {author}
  {\bibfnamefont {N.~P.}\ \bibnamefont {Ong}}, \ and\ \bibinfo {author}
  {\bibfnamefont {R.~J.}\ \bibnamefont {Cava}},\ }\href {\doibase
  10.1038/nphys360} {\bibfield  {journal} {\bibinfo  {journal} {Nature
  Physics}\ }\textbf {\bibinfo {volume} {2}},\ \bibinfo {pages} {544} (\bibinfo
  {year} {2006})}\BibitemShut {NoStop}%
\bibitem [{\citenamefont {Shiina}\ \emph {et~al.}(1998)\citenamefont {Shiina},
  \citenamefont {Sakai}, \citenamefont {Shiba},\ and\ \citenamefont
  {Thalmeier}}]{Shiina1998}%
  \BibitemOpen
  \bibfield  {author} {\bibinfo {author} {\bibfnamefont {R.}~\bibnamefont
  {Shiina}}, \bibinfo {author} {\bibfnamefont {O.}~\bibnamefont {Sakai}},
  \bibinfo {author} {\bibfnamefont {H.}~\bibnamefont {Shiba}}, \ and\ \bibinfo
  {author} {\bibfnamefont {P.}~\bibnamefont {Thalmeier}},\ }\href {\doibase
  10.1143/JPSJ.67.941} {\bibfield  {journal} {\bibinfo  {journal} {Journal of
  the Physical Society of Japan}\ }\textbf {\bibinfo {volume} {67}},\ \bibinfo
  {pages} {941} (\bibinfo {year} {1998})},\ \Eprint
  {http://arxiv.org/abs/https://doi.org/10.1143/JPSJ.67.941}
  {https://doi.org/10.1143/JPSJ.67.941} \BibitemShut {NoStop}%
\bibitem [{\citenamefont {Chen}\ \emph {et~al.}(2010)\citenamefont {Chen},
  \citenamefont {Pereira},\ and\ \citenamefont {Balents}}]{ChenBalents2010}%
  \BibitemOpen
  \bibfield  {author} {\bibinfo {author} {\bibfnamefont {G.}~\bibnamefont
  {Chen}}, \bibinfo {author} {\bibfnamefont {R.}~\bibnamefont {Pereira}}, \
  and\ \bibinfo {author} {\bibfnamefont {L.}~\bibnamefont {Balents}},\ }\href
  {\doibase 10.1103/PhysRevB.82.174440} {\bibfield  {journal} {\bibinfo
  {journal} {Phys. Rev. B}\ }\textbf {\bibinfo {volume} {82}},\ \bibinfo
  {pages} {174440} (\bibinfo {year} {2010})}\BibitemShut {NoStop}%
\end{thebibliography}%

\appendix

\section{Tight-binding model}
The bilayer model for KTaO$_3$ (KTO) oriented along the (111) direction consists of three orbitals and two layers \cite{xiao}. Including the spin-orbit coupling, $\lambda$, the secular matrix is of order 12. We use the basis $\left\lbrace\ket{d_{1,yz\uparrow}}\right.$, $\ket{d_{1,xz\uparrow}}$, $\ket{d_{1,xy\uparrow}}$, $\ket{d_{2,yz\uparrow}}$, $\ket{d_{2,xz\uparrow}}$, $\ket{d_{2,xy\uparrow}}$, $\ket{d_{1,yz\downarrow}}$, $\ket{d_{1,xz\downarrow}}$, $\ket{d_{1,xy\downarrow}}$, $\ket{d_{2,yz\downarrow}}$, $\ket{d_{2,xz\downarrow}}$,  $\left. \ket{d_{2,xy\downarrow}} \right\rbrace$, where indices $1, 2$ label the layers. For purposes here, we only consider the largest hopping term, $t$, which is diagonal in the orbital index, but off-diagonal in the layer index. The secular matrix has the following form:
\begin{widetext}
\begin{equation}
\begin{split}
\hat{\mathcal{H}}(\bm{k})=
 &\left( 
 {\begin{array}{cccccccccccc}
     0  & \frac{\Delta}{2}+\frac{i\lambda}{2} & \frac{\Delta}{2} &   \xi_1 & 0 & 0 & 0 & 0 & -\frac{\lambda}{2} & 0 & 0 & 0 \\[\smallskipamount]
    \frac{\Delta}{2}-\frac{i\lambda}{2} &  0  & \frac{\Delta}{2} & 0 &   \xi_2 & 0 & 0 & 0 & \frac{i\lambda}{2} & 0 & 0 & 0 \\[\smallskipamount]
    \frac{\Delta}{2} & \frac{\Delta}{2} &  0  & 0 & 0 & \xi_3 & \frac{\lambda}{2} & -\frac{i\lambda}{2} & 0 & 0 & 0 & 0 \\[\smallskipamount]
    \xi_1^* & 0 & 0 &  0  & \frac{\Delta}{2}+\frac{i\lambda}{2} & \frac{\Delta}{2} & 0 & 0 & 0 & 0 & 0 & -\frac{\lambda}{2} \\[\smallskipamount] 
    0 &   \xi_2^* & 0 & \frac{\Delta}{2}-\frac{i\lambda}{2} &  0  & \frac{\Delta}{2} & 0 & 0 & 0 & 0 & 0 & \frac{i\lambda}{2} \\[\smallskipamount]
    0 & 0 & \xi_3^* & \frac{\Delta}{2} & \frac{\Delta}{2} &  0  & 0 & 0 & 0 & \frac{\lambda}{2} & -\frac{i\lambda}{2} & 0 \\[\smallskipamount]
    0 & 0 & \frac{\lambda}{2} & 0 & 0 & 0 &  0  & \frac{\Delta}{2}-\frac{i\lambda}{2} & \frac{\Delta}{2} &   \xi_1 & 0 & 0 \\[\smallskipamount]
    0 & 0 & \frac{i\lambda}{2} & 0 & 0 & 0 & \frac{\Delta}{2}+\frac{i\lambda}{2} &  0  & \frac{\Delta}{2} & 0 &   \xi_2 & 0 \\[\smallskipamount]
    -\frac{\lambda}{2} & -\frac{i\lambda}{2} & 0 & 0 & 0 & 0 & \frac{\Delta}{2} & \frac{\Delta}{2} &  0  & 0 & 0 & \xi_3 \\[\smallskipamount]
    0 & 0 & 0 & 0 & 0 & \frac{\lambda}{2} &   \xi_1^* & 0 & 0 &  0  & \frac{\Delta}{2}-\frac{i\lambda}{2} & \frac{\Delta}{2} \\[\smallskipamount]
    0 & 0 & 0 & 0 & 0 & \frac{i\lambda}{2} & 0 &   \xi_2^* & 0 & \frac{\Delta}{2}+\frac{i\lambda}{2} &  0  & \frac{\Delta}{2} \\[\smallskipamount]
    0 & 0 & 0 & -\frac{\lambda}{2} & -\frac{i\lambda}{2} & 0 & 0 & 0 & \xi_3^* & \frac{\Delta}{2} & \frac{\Delta}{2} & 0   
\end{array}} \right).
\end{split}
 \label{eq:Matrix_12x12}
\end{equation}
\end{widetext}
where we define:
\begin{equation}
\begin{split}
 &\xi_1=-te^{ik_2 c}\left[1+e^{i\left(\frac{\sqrt{3}k_1 c}{2}-\frac{3k_2 c}{2}\right)}\right]\\
 &\xi_2=-te^{ik_2 c}\left[1+e^{-i\left(\frac{\sqrt{3}k_1 c}{2}+\frac{3k_2 c}{2}\right)}\right]\\
 &\xi_3=-2t\cos\left(\frac{\sqrt{3}k_1 c}{2}\right)e^{-i\frac{k_2 c}{2}}.
 \label{eq:SM1}
\end{split}
\end{equation}
\begin{figure}
\includegraphics[width=0.49\columnwidth]{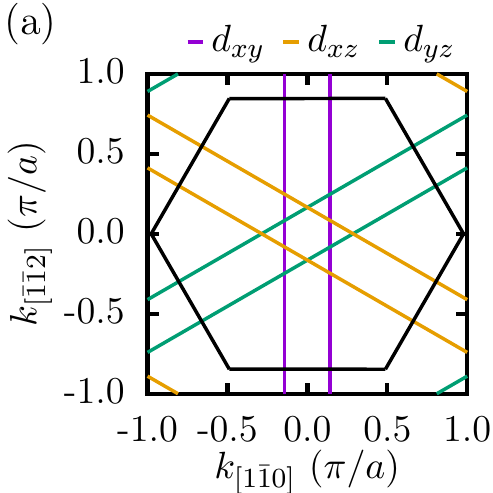}
\includegraphics[width=0.49\columnwidth]{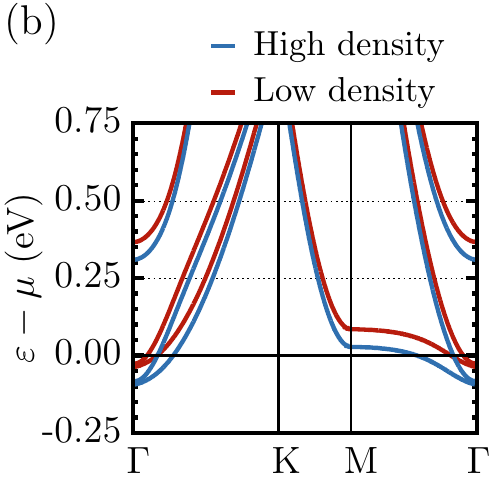}
\includegraphics[width=0.98\columnwidth]{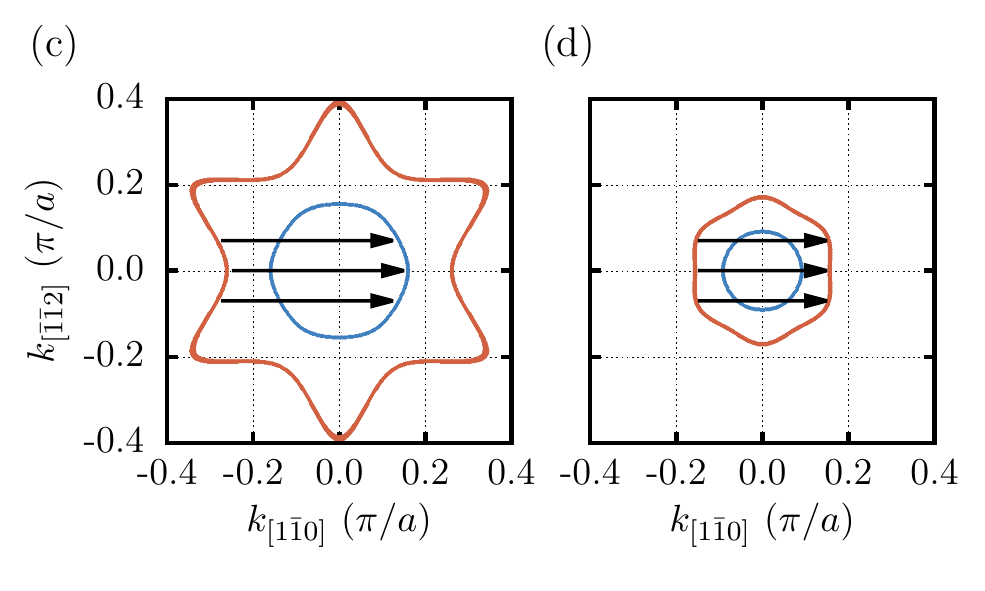}
\caption{(a) Fermi surface of (111) KTO in a $t$-only bilayer model, with $\Gamma$-$K$ along the
horizontal axis and $\Gamma$-$M$ along the vertical axis.  The bands are labeled by their $t_{2g}$ content, and
the hexagon marks the surface Brillouin zone boundary. (b) Electronic band structure
including spin-orbit coupling and a small trigonal distortion.
The two sets of curves correspond to two chemical potentials adjusted to match the
carrier densities reported by Bruno {\it et al.}~\cite{bruno} (high density) and Liu {\it et al.}~\cite{liu} (low density).
Fermi surface for (c) the high density case and (d) the low density case.
The horizontal arrows indicate the nestings along $\Gamma$-$K$ that were identified by the susceptibility.}
\label{fig1}
\end{figure}
Here, $k_1 \!\equiv\! k_{[1\bar{1}0]}$ and $k_2\! \equiv\! k_{[11\bar{2}]}$ refer to orthogonal momenta in the hexagonal
surface Brillouin zone (BZ), with $c$ being the projection of the bulk lattice constant, $a$,
onto the (111) plane (i.e., $c=\sqrt{2/3}a$ with $a=3.99$~\AA{}). The in-plane (111) lattice constant is $\sqrt{3}c$, which sets the scale for
the surface BZ. We chose $t=-1$~eV which, in the absence of other terms, fixes the total bandwidth of the model to $4|t|$ (the value
of $t$ being set by the bulk KTO bandwidth).
The SOC term for $t_{2g}$ orbitals has a value of 265 meV, leading to a bare quartet-singlet splitting
of 397.5 meV at $\Gamma$ (i.e., when $t$ is turned off). This value has been chosen to match
the large spin-orbit splitting reported from ab initio calculations of KTO~\cite{bruno}.
In addition, a small on-site trigonal distortion ($\Delta=10$~meV) is included that is off-diagonal in the orbital index.
With the other terms turned off, this leads to a 15 meV splitting at $\Gamma$. 
We have found that the other (smaller) hoppings, as well as the potential difference between the two layers,
which were considered in Ref.~\onlinecite{xiao}, play only a minor role and are not included here.

The $t$-only model generates a Fermi surface consisting of three pairs of parallel lines along each the $\Gamma$-$M$ direction (Fig.~\ref{fig1}a), which are perfectly nested along $\Gamma$-$K$. Once spin-orbit coupling is included, the band structure changes and the flat band in the $t$-only model becomes dispersive as seen in Fig.~\ref{fig1}b. With this model, we have performed calculations at two densities (set by the chemical potential), a higher one to match the ARPES results~\cite{bruno}, and a lower one to match that of Liu {\it et al.}~\cite{liu}.  These two densities give rise to different Fermi surfaces (Figs.~\ref{fig1}c and ~\ref{fig1}d). For the higher density, the outer pocket has a star-like shape and the inner pocket has a hexagonal shape. In this case, the maximum in the susceptibility does not come along $\Gamma$-$M$ (due to nesting of the outer star surface), but rather along $\Gamma$-$K$ due to nesting between the outer and inner surfaces. However, in the lower density case, the outer surface becomes hexagonal, and the maximum in the susceptibility is due to nesting along $\Gamma$-$K$ for the outer surface, as shown in the main text.

\section{Rashba Term}
 
The Rashba term can be derived by generalizing the work of Khalsa {\it et al.}~\cite{khalsa} to the (111) case.
As (111) is parallel to $x+y+z$, more terms enter than in their (001) case.  As an example, the $yz$ to $xy$ hopping
along the Ta-O-Ta cubic $z$ direction is of the form $\bra{yz}y\rangle\bra{y}E_x\ket{xy}$ where the first matrix element is 
$+t_{pd}$, with $t_{pd}$ the overlap integral between Ta $5d$ and O $2p$ orbitals,
and the second one is the inversion breaking term due to the electric field along the $x$ direction, $E_x$.
This results in the following matrix elements, $\gamma$, to be added to the secular matrix:
\begin{equation}
\begin{split}
&\gamma_{1 yz, 2 xz} = 2it_R\sin\left(\frac{\sqrt{3}k_1 c}{2}\right)e^{-i\frac{k_2 c}{2}} \\
&\gamma_{1yz, 2 xy} = t_Re^{ik_2 c}\left[1-e^{-i\left(\frac{\sqrt{3}k_1 c}{2}+\frac{3k_2 c}{2}\right)}\right]  \\
&\gamma_{1 xz, 2 xy} = t_Re^{ik_2 c}\left[1-e^{i\left(\frac{\sqrt{3}k_1 c}{2}-\frac{3k_2 c}{2}\right)}\right] 
\label{eq:SM2}
\end{split}
\end{equation}
where diagonality in the spin index is implicit. (Note that this corresponds to an orbital inversion symmetry breaking and
is thus spin-diagonal.) The spin-splitting arises from the combination of this inversion breaking term and the atomic SOC.
We have found that a value of $t_R$ of 2 meV is needed to reproduce the suggested Rashba splitting along
$\Gamma$-$K$ of Bruno {\it et al.}~\cite{bruno}.  This value leads to an almost uniform splitting of around 0.005$\pi/c$
for the outer surface of the lower density case of Liu {\it et al.}~\cite{liu}. 

\section{j=3/2 effective model and multipoles}

\newcommand{\be}{\begin{equation}}
\newcommand{\ee}{\end{equation}}
\newcommand{\bea}{\begin{eqnarray}}
\newcommand{\eea}{\end{eqnarray}}
\newcommand{\bJ}{\bf{J}}

For a cubic crystal, the symmetry-allowed continuum 4-band Luttinger model for the 3D bulk dispersion
near the $\Gamma$-point is given, to ${\cal O}(k^2)$, by
\be
\!\! H^{\rm Lutt}_{\rm 3D} \!=\! \alpha_1 k^2\! \hJ_0 + \alpha_2 (\bk \cdot \hat{\bJ})^2 +  \alpha_3\!\! \sum_{i=x,y,z} k^2_i \! \hJ_i^2 \!
\ee 
Here, $\hJ_i$ refer to spin-$3/2$ angular momentum operators (with $i\!=\! x,y,z$), $\hJ_0$ is the $4\! \times\! 4$ identity matrix, and we measure
momenta in units of $1/a$ where $a\!\approx\!4$\AA~ is the cubic lattice constant.
For KTaO$_3$, we find that a single parameter model, with $\alpha_1\!=\!\alpha_2\!=\!0$ and $\alpha_3\!=\!0.2$\,eV,
captures the band dispersion near the $\Gamma$ point. We impose a momentum cutoff $\Lambda = \pi/3$.

\begin{figure}
	\centering
	\includegraphics[width=\linewidth]{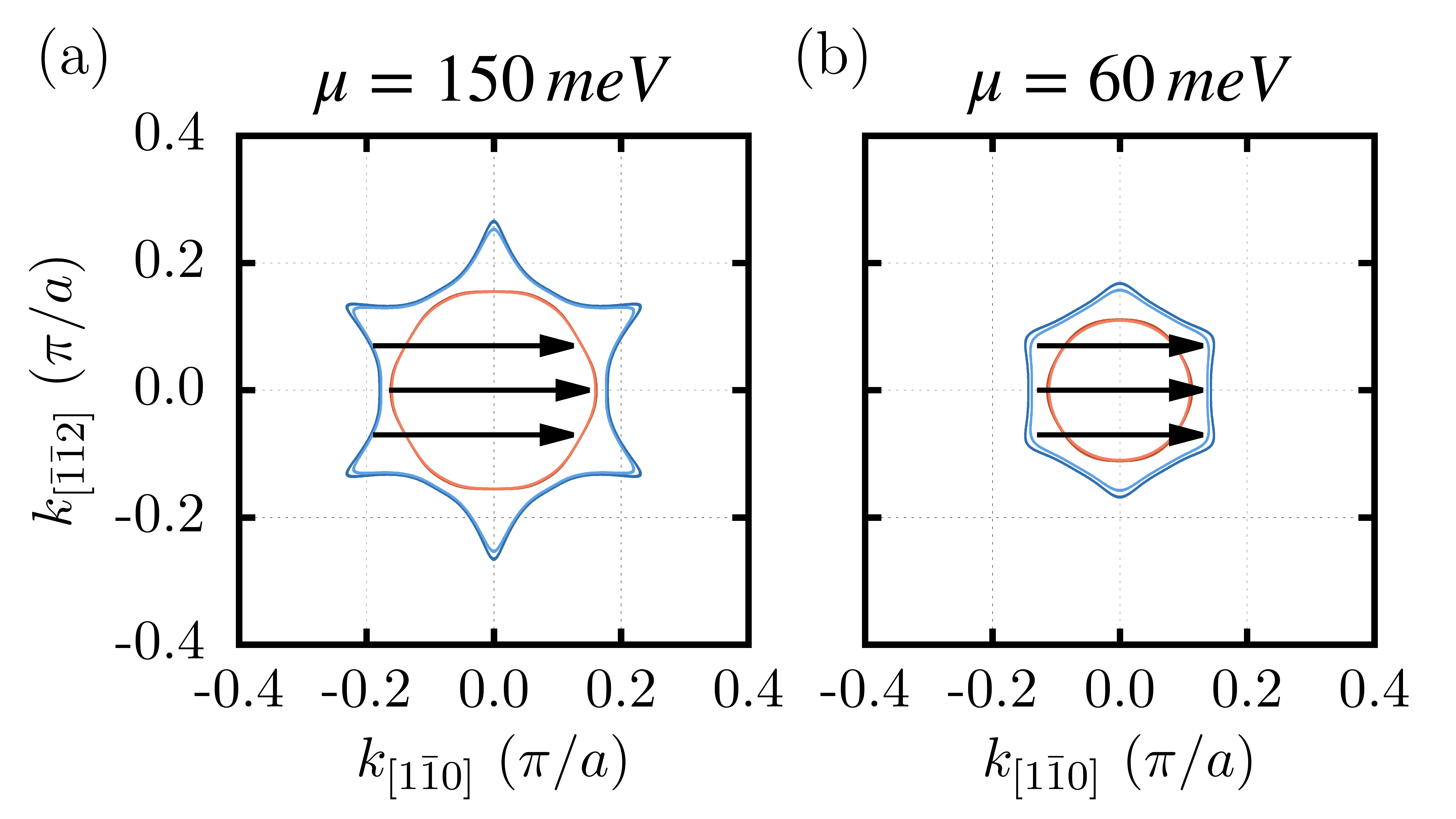}
	\caption{Fermi surfaces for the 2DEG upon including the Rashba SOC and trigonal distortion, shown for chemical potential 
	$\mu=150$\,meV ($n \approx 6.5 \times 10^{13}$/cm$^2$) and $\mu=60$\,meV 
	($n \approx 3.5 \times 10^{13}$/cm$^2$).
	In each case, the spin-split FSs are more clearly visible for the outer pair of bands, with the splitting being slightly more
	significant along the $k_2 \equiv k_{[\bar{1}\bar{1}2]}$ direction which corresponds to the $\Gamma$-$M$ direction.}
	\label{fig:FSjmodel}
\end{figure}

To describe the $(111)$ 2DEG, we
take this dispersion and project it to 2D, expressing it in terms of orthogonal momentum components in the
plane of the 2DEG, namely
$k_1$ and $k_2$ which are respectively along the $(1\bar{1}0)$ and $(11\bar{2})$ 
directions, so
$k_x \!=\!  (k_2 + \sqrt{3} k_1)/\sqrt{6}$, $k_y\! =\! (k_2-\sqrt{3} k_1)/\sqrt{6}$, and $k_z\!=\! -2 k_2/\sqrt{6}$. The projected
2D Hamiltonian is
\bea
\!\!\!\! H^{(0)}_{2D} &\!\!=\!& \alpha_3\! \! \left[ \! \frac{(\sqrt{3} k_1 \!+\! k_2)^2}{6} \hJ^2_x  \!+\!  \frac{(\sqrt{3} k_1 \!-\! k_2)^2}{6} \hJ^2_y \!+\! \frac{2 k^2_2}{3} \hJ^2_z \!\right]
\eea
Going beyond these terms which descend from the bulk dispersion, we need to incorporate additional symmetry allowed terms in order
to describe the 2DEG dispersion. We begin by considering mirror symmetry $M_1$, time reversal ${\cal T}$, three-fold rotation ${\cal R}_{2\pi/3}$, and 
inversion ${\cal I}$, and then incorporate Rashba terms from  breaking ${\cal I}$. For convenience, we define $k_\pm = k_1 \pm i k_2$.
Under lattice symmetry operations, the momenta and $\vec J$ transform as:
\begin{eqnarray}
M_1&:& \left\{ \begin{array}{lcl} (k_1,k_2) & \rightarrow & (-k_1,k_2) \\ 
(J_x,J_y,J_z) & \rightarrow & (-J_y,-J_x,-J_z)
         \end{array} \right. \nonumber \\
R_{2\pi/3}&:& \left\{ \begin{array}{lcl} k_\pm & \rightarrow
    & k_\pm e^{\pm i 2\pi/3} \\
(J_x,J_y,J_z) & \rightarrow & (J_y,J_z,J_x)
\end{array} \right. \nonumber \\
{\cal I} &:& \left\{ \begin{array}{lcl}  (k_1,k_2) & \rightarrow
    & (-k_1,-k_2) \\
(J_x,J_y,J_z) & \rightarrow & (J_x,J_y,J_z)
\end{array}\right.
\end{eqnarray}
If inversion and time-reversal are unbroken, cubic terms in the momenta are ruled out, and the next important terms which we find 
capture the hexagonal shape of the 2DEG dispersion are sixth order terms in momenta,
\bea
H^{(1)}_{2D} &\!=\!& \left[\beta_1 (k_+^6 \!+\! k_-^6) \!+\! \beta_2 (k_+^3\!+\! k_-^3)^2\right] \hJ_0 \notag\\
&+& \beta_3 (k_+^3 \!-\! k_-^3)^2 \hJ_3^2
\eea
where $\hJ_3=(\hJ_x+\hJ_y+\hJ_z)/\sqrt{3}$.
Finally, we incorporate two weaker 
terms: an effective trigonal distortion $\tilde{\Delta}$, and a Rashba coupling $\tilde{\gamma}$ from inversion breaking, via
\bea
\!H^{(2)}_{2D} &\!=\!&\tilde{\Delta} \hJ_3^2  + \tilde{\gamma} (\hJ_1 k_2 \!-\! \hJ_2 k_1)
\eea
where $J_1 \!=\! (J_x\!-\! J_y)/\sqrt{2}$ and $J_2 \!=\! (J_x\!+\!J_y\!-\!2 J_z)/\sqrt{6}$. We set $(\beta_1,\beta_2,\beta_3)\!=\! (0.35,0.6,-0.65)$\,eV,
and fix the weaker terms to be $(\tilde{\Delta},\tilde{\gamma}) \!=\! (7,7)$\,meV in order to match the tight-binding results for the
spin splitting of the FSs and the splitting of the $j=3/2$ quartet at the $\Gamma$-point.  This Hamiltonian $H_{\rm 2DEG} \!= \!H^{(0)}_{2D}+\!H^{(1)}_{2D}+\!H^{(2)}_{2D}$
 may be viewed as a simple continuum $j=3/2$ model for the KTaO$_3$ (111) 2DEG for
low to moderate dopings. The Fermi surfaces from this $k\cdot p$ Hamiltonian are shown in Fig.~\ref{fig:FSjmodel}, and are in reasonable agreement
with the FSs from the tight-binding model discussed in the main text.

The multipole moments within the $j=3/2$ quartet are given in the table below. We have normalized these multipole operators such that
${\rm Tr}({\cal M}_\alpha^2)=4 j(j+1)/3$. This normalization is chosen such that the dipole operator matrices are simply the usual $\hJ_i$ matrices.
Different normalization schemes, as in Refs.~\onlinecite{Shiina1998,ChenBalents2010}, would lead to quantitative
changes in the susceptibility results in the main text, but the key conclusions, that the dominant eigenvalue of $\chi$ stems from interband (12+21) nesting,
and that it involves dipoles and octupoles, remains robust.

\begin{table}[h]
\centering 
\begin{tabular}{|l|c|c|} 
\hline
Multipole  & Symmetry & Operator  \\ 
\hline \hline 
Dipole  & $\Gamma_4$ & ${\cal M}_x = J_x$  \\  
        &            & ${\cal M}_y = J_y$ \\
        &            & ${\cal M}_z = J_z$ \\
        \hline
Quadrupole 
&             $\Gamma_5$ & ${\cal M}_{yz} = \sqrt{\frac{5}{12}} \overline{J_y J_z} $ \\
&&                        ${\cal M}_{xz} = \sqrt{\frac{5}{12}} \overline{J_x J_z}  $    \\
&&                       ${\cal M}_{xy} = \sqrt{\frac{5}{12}} \overline{J_x J_y}  $  \\
& $\Gamma_3$ &    ${\cal M}_{x^2-y^2} = \sqrt{\frac{5}{12}} (J^2_x- J^2_y) $   \\
&&                   ${\cal M}_{3z^2} = \sqrt{\frac{5}{36}} (3 J_z^2- J(J+1)) $   \\ 
\hline                              
Octupole & $\Gamma_2$ &   $ {\cal M}_{xyz} = \sqrt{\frac{5}{27}} \overline{J_x J_y J_z}  $ \\
          & $\Gamma_4$ &   $ {\cal M}_x^{\alpha} = \frac{2}{3} (J_x^3 - \frac{1}{2}(\overline{J_x J_y^2} + \overline{J_x J_z^2}))$ \\
          &&               $ {\cal M}_y^{\alpha} = \frac{2}{3} (J_y^3 - \frac{1}{2}(\overline{J_y J_x^2} + \overline{J_y J_z^2}))$ \\
          &&               $ {\cal M}_z^{\alpha} = \frac{2}{3} ( J_z^3 - \frac{1}{2}(\overline{J_z J_x^2} + \overline{J_z J_y^2}))$ \\
          & $\Gamma_5$ &   $ {\cal M}_x^{\beta} =  \sqrt{\frac{5}{27}} \overline{J_x (J_y^2-J_z^2)} $ \\
          &&               $ {\cal M}_y^{\beta} = \sqrt{\frac{5}{27}} \overline{J_y (J_z^2-J_x^2)} $ \\
          &&               $ {\cal M}_z^{\beta} = \sqrt{\frac{5}{27}}\overline{J_z (J_x^2-J_y^2)} $\\
\hline
\end{tabular}
\caption{The overline symbol indicates a sum over all the possible permutations of the operators,
$\overline{J_x J_z^2} = J_x J_z^2 + J_z J_x J_z + J_z^2 J_x$. This table is adapted from Refs.~\onlinecite{Shiina1998, ChenBalents2010} but with
modified normalization.}
\label{table1}
\end{table}

In order to compare these results with the results from the tight-binding model calculations, we also plot in Fig.~\ref{fig4} the susceptibility eigenvalues for intraband and
interband orders when we truncate the susceptibility matrix to just the dipole operators, where we find qualitative agreement with the results from Fig.~4 of the main text.
\begin{figure}
\includegraphics[width=0.95\columnwidth]{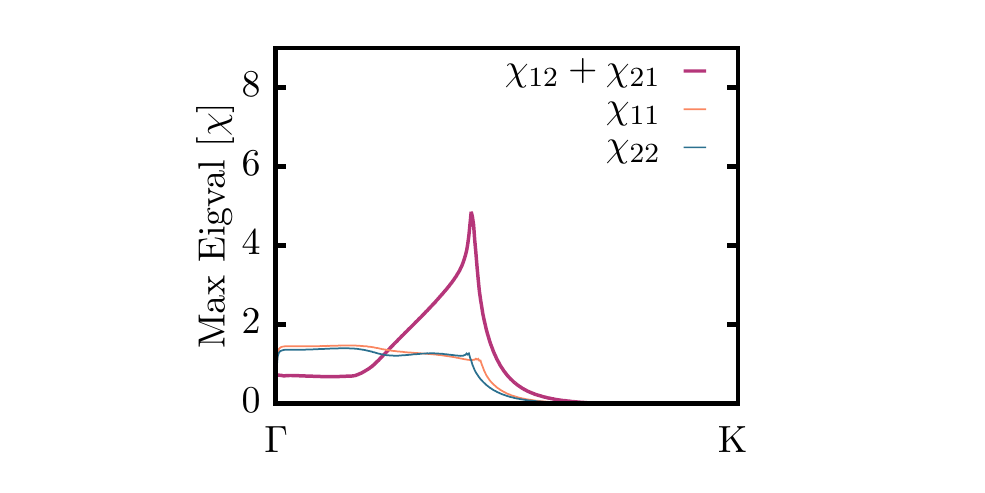}
\caption{Largest eigenvalues of the $3\times 3$ susceptibility matrix truncated to just the dipole operators plotted as a function of momentum along $\Gamma$-$K$.}
\label{fig4}
\end{figure}

Fig.~\ref{fig3}a-f shows the real and imaginary components of the eigenvectors corresponding to the peaks in $\chi$ shown in the main text
where we diagonalize the full $15\times 15$ susceptibility matrix.
\begin{figure}
\includegraphics[width=0.95\columnwidth]{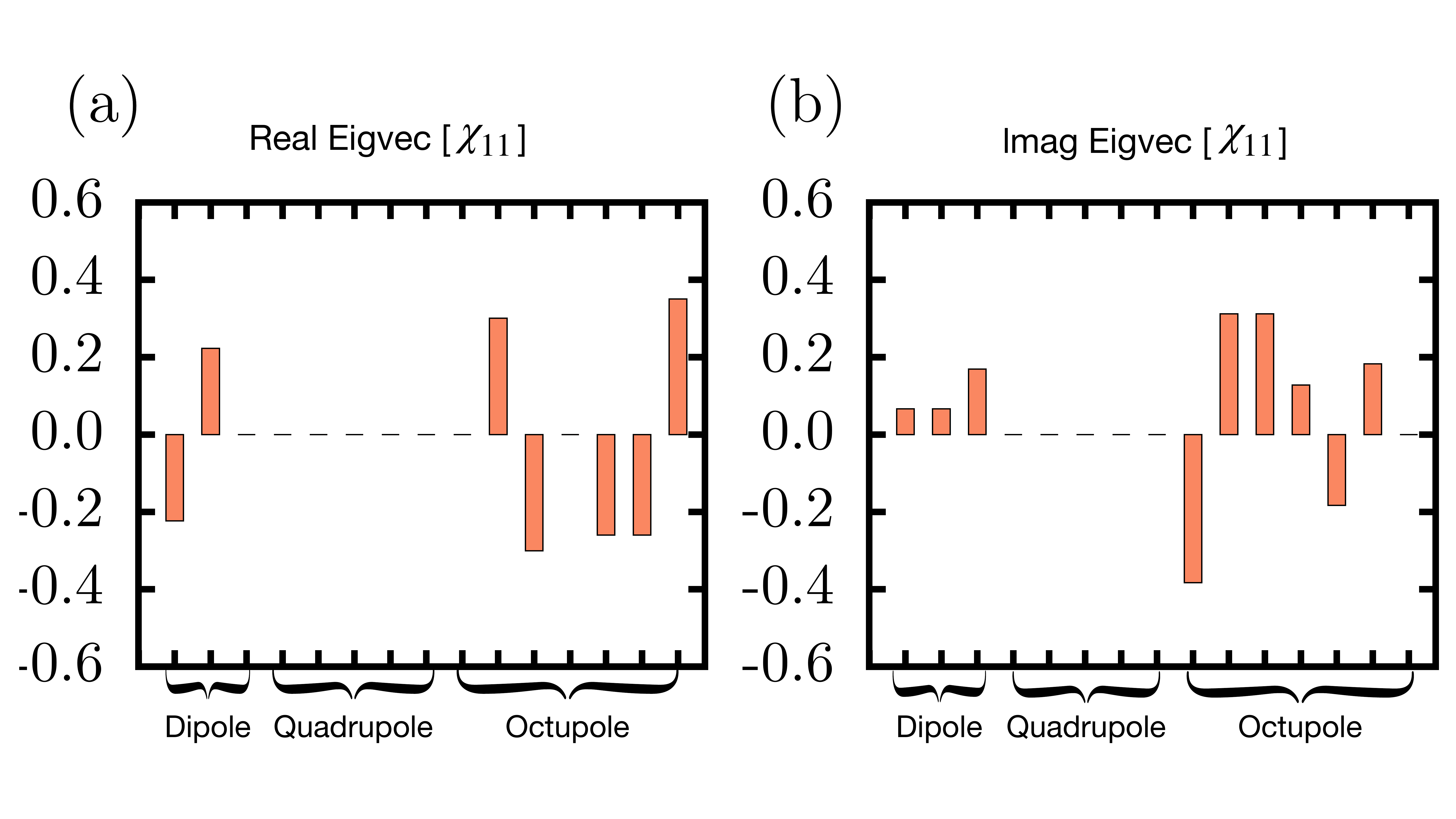}
\includegraphics[width=0.95\columnwidth]{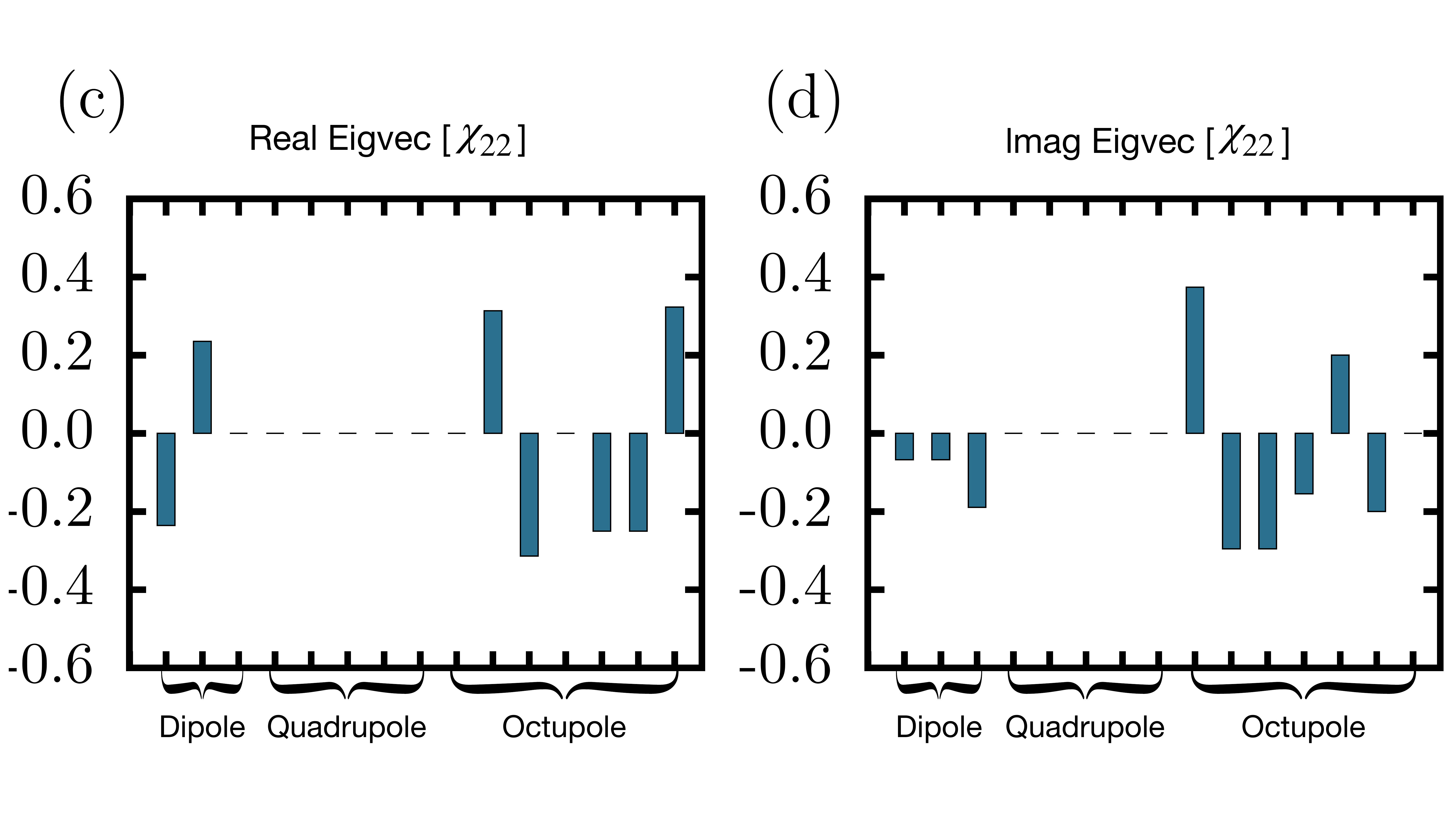}
\includegraphics[width=0.95\columnwidth]{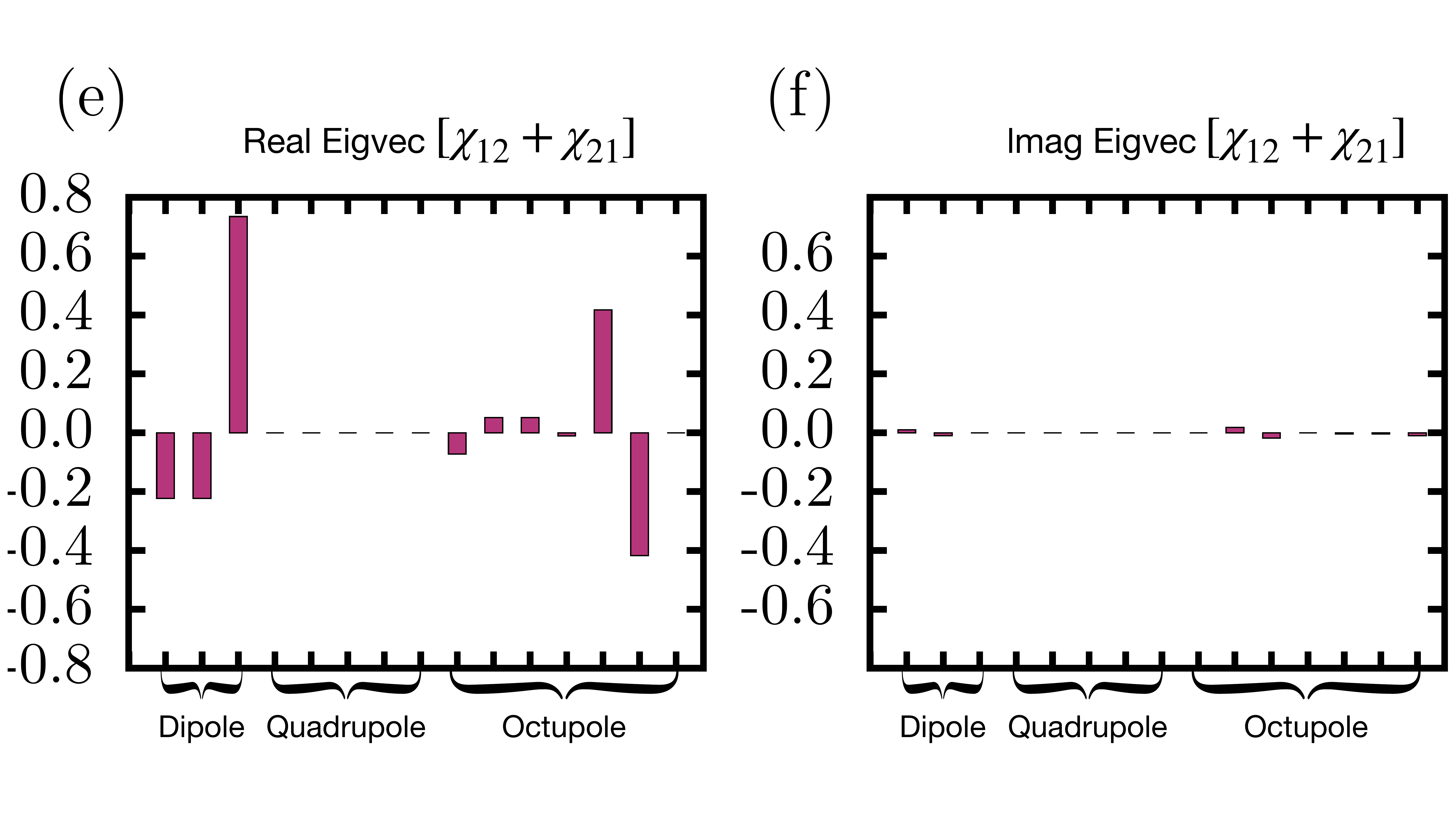}
\caption{Real and imaginary components of the dominant eigenvector corresponding to the peaks in $\chi$ for (a,b) intraband $\chi_{11}$,
(c,d) intraband $\chi_{22}$, and (e,f) interband $\chi_{12}+\chi_{21}$. For the leading instability which corresponds to an interband order, the symmetry
breaking pattern involves time-reversal breaking dipolar and octupolar modulations.}
\label{fig3}
\end{figure}

\end{document}